\documentclass[acmsmall,screen,nonacm]{acmart}
\AtBeginDocument{%
  }
\usepackage[table,dvipsnames]{xcolor}
\usepackage[most]{tcolorbox}
\usepackage{enumitem}
\definecolor{keyfindingsblue}{RGB}{95,158,202}
\definecolor{keyfindingsbg}{RGB}{235,245,252}
\newtcolorbox{keyfindings}[2][]{
  enhanced,
  breakable,
  colback=keyfindingsbg,
  colframe=keyfindingsblue,
  colbacktitle=keyfindingsblue,
  coltitle=white,
  fonttitle=\bfseries,
  fontupper=\small,
  title={#2 -- Key Findings},
  boxrule=0.5pt,
  arc=2pt,
  left=4pt,
  right=4pt,
  top=3pt,
  bottom=3pt,
  toptitle=2pt,
  bottomtitle=2pt,
  before skip=8pt,
  after skip=10pt,
  #1
}
\newenvironment{keyfindingslist}
  {\begin{itemize}[leftmargin=1.25em,labelsep=0.45em,topsep=1pt,itemsep=1pt,parsep=0pt,partopsep=0pt]}
  {\end{itemize}}
\usepackage{multirow}
\usepackage{subfig}
\usepackage{placeins}
\usepackage{needspace}
\usepackage{tabularx}
\usepackage{longtable}
\usepackage{xurl}
\definecolor{benchmarkruntime}{RGB}{223,249,255}
\definecolor{benchmarkconfig}{RGB}{239,234,250}
\definecolor{benchmarkexampleframe}{RGB}{105,116,128}
\definecolor{benchmarkexamplecode}{RGB}{247,248,250}

\begin{document}

\title[Model Context Protocol Software Repair]{Repairing Model Context Protocol Software: Fix Patterns, Benchmark, and Agent Evaluation}

\author{Mina Taraghi}
\email{mina.taraghi@polymtl.ca}
\orcid{0009-0007-8250-4200}
\affiliation{%
  \institution{Polytechnique Montreal}
  \city{Montreal}
  \country{Canada}
}

\author{Mohammad Mehdi Morovati}
\email{mehdi.morovati@polymtl.ca}
\orcid{0000-0002-7942-4791}
\affiliation{%
  \institution{Polytechnique Montreal}
  \city{Montreal}
  \country{Canada}
}

\author{Foutse Khomh}
\email{foutse.khomh@polymtl.ca}
\orcid{0000-0002-5704-4173}
\affiliation{%
  \institution{Polytechnique Montreal}
  \city{Montreal}
  \country{Canada}
}

\begin{abstract}
Model Context Protocol (MCP) servers connect language models to external tools and services. Prior work has characterized where MCP servers fail, but less is known about how developers repair them or how well automated systems can perform these repairs. We analyze 221 issues with verified fixes and classify 197 of those repairs into 10 repair groups and 60 action patterns using independent open coding and affinity diagramming. Fixes change a median of two files and 59 lines. Patch structure differs across repair groups, but we find no significant differences in repair timing. From a separate corpus of bug reports, we construct MCP-RepairBench, a benchmark of 60 real repair tasks from 25 repositories with pinned buggy and fixed revisions, behavioral evaluators, and reproducible offline environments. An MCP-aware repair agent with evaluator feedback resolves 34 tasks on the first attempt and 41 within three attempts, compared with 18 and 26 for a baseline using the same model and per-attempt limits. Remaining failures often involve configuration, packaging, exact response requirements, or changes that must be coordinated across files.
\end{abstract}

\ccsdesc[500]{Software and its engineering~Software testing and debugging}
\ccsdesc[300]{Software and its engineering~Software maintenance tools}

\keywords{Model Context Protocol, software testing, bug-fix patterns, automated program repair, repair benchmarks}

\maketitle

\section{Introduction}

\noindent Large language models (LLMs) are increasingly embedded in agentic systems that extend their capabilities through external tools, APIs, and services~\cite{minaeeLLMSurvey2024,liAPIBank2023}. The Model Context Protocol (MCP) standardizes these interactions, allowing hosts to discover and invoke tools, access resources, and exchange contextual information with MCP servers~\cite{mcpSpec2025,houMCP2025}. MCP servers therefore form an integration layer between LLM-based agents and the external software systems they operate. Their correctness depends not only on application logic, but also on the consistency of contracts spanning tool interfaces, protocol messages, configuration, packaging, transport, and runtime behavior.

\noindent These cross-layer dependencies create distinctive reliability challenges. An MCP server can correctly implement its intended domain operation and still fail in practice: its advertised tool schema may disagree with the implementation that processes requests, its launch configuration may omit a required value, its package may lack a runtime resource, or logs and error messages written to \texttt{stdout} may interfere with protocol messages. Such failures are particularly relevant in a rapidly evolving and heterogeneous ecosystem, where MCP servers vary in implementation, dependencies, deployment environments, and client connectivity~\cite{mcpMeasurement2025}. Prior work shows that real MCP faults frequently arise at server--host, server--tool, resource, runtime, and protocol-configuration boundaries~\cite{ourPaper}. These findings establish where MCP systems fail, but they leave an important question unanswered: \emph{how are such failures actually repaired?}

\noindent Repairing an MCP failure requires more than locating a defective line of code. Developers must identify the artifacts involved, determine which contract or assumption has been violated, formulate changes that address the underlying cause, and verify that the resulting system exhibits the required behavior. The resulting patch may be localized to one artifact or may require coordinated changes across several artifacts. Prior research on developer-written patches has shown that recurring repair operations can reveal systematic maintenance patterns and support repair-oriented abstractions and automated techniques~\cite{panFixPatterns2009,camposFixPatterns2019,fixminer2020}. Yet for MCP, we do not know whether comparable regularities exist, what forms they take, or how they relate to the integration boundaries at which faults arise. This missing repair perspective limits our understanding of MCP maintenance and, more importantly, leaves unclear what an automated repair system would need to change and validate.

\noindent These questions are directly relevant to automated repair. Recurring repair operations may provide useful structure for automation, whereas repairs that depend on interactions among protocol, configuration, packaging, and runtime behavior may require broader system-level reasoning. Developer fixes can therefore inform both the capabilities that automated systems should provide and the tasks on which those capabilities should be evaluated. Such evaluation must go beyond whether a generated patch passes an available test suite. A patch can satisfy incomplete tests while failing to restore the behavior required by the original issue, a longstanding concern in automated program repair~\cite{qi2015,gazzola2019}. Meaningful evaluation consequently requires real faults, the repository states in which they occurred, and behavioral validation that distinguishes genuine repairs from narrow workarounds.

\noindent We address these gaps through a staged empirical study. First, using a previously validated MCP fault dataset, we identify 221 issues for which a developer fix is available. Through independent open coding and affinity diagramming, we characterize the repair operations in these fixes and derive 10 repair groups and 60 action patterns for the 197 repairs that can be localized to a sufficiently specific operation. We examine the artifacts involved, the structure of the resulting patches, and how repair characteristics vary across groups. Second, using a separate extended bug-report corpus, we construct MCP-RepairBench from real fix-linked cases. Each task pairs pinned buggy and developer-fixed repository revisions with a case-specific behavioral evaluator, a reproducible offline environment, and preservation checks where applicable. The separate corpus provides independent cases for automated-repair evaluation rather than reusing the issues from which the repair taxonomy was derived. The benchmark follows the real-fault, repository-level tradition of automated repair~\cite{defects4j2014,bugsinpy2020,swebench2024}, while emphasizing behavioral validation because passing an incomplete test suite does not necessarily establish that the reported problem has been resolved~\cite{swebenchCorrectness2026}.

\noindent Finally, we use MCP-RepairBench to evaluate an MCP-aware repair agent that takes an issue description and buggy repository as input and generates a repair using evaluator feedback. We measure task resolution and analyze execution traces, resource use, and failure mechanisms to identify which repair challenges remain difficult for automated systems. The study thus connects empirical understanding of developer repairs with reproducible evaluation of automated repair.

\noindent We organize the study around the following research questions (RQ):
\begin{itemize}
\item[\textbf{RQ1.}] What are the recurring repair groups and action patterns in fixes for MCP bugs?
\item[\textbf{RQ2.}] What are the characteristics of MCP bug fixes, and how do they vary across repair groups?
\item[\textbf{RQ3.}] How can real MCP bug reports and developer fixes be curated into a reproducible benchmark for automated repair?
\item[\textbf{RQ4.}] To what extent can an MCP-aware repair agent resolve real MCP bug-repair tasks, and what challenges remain?
\end{itemize}

\noindent This paper makes three main contributions:
\begin{itemize}
\item \textbf{Empirical characterization of MCP repair.} We provide an independently coded, two-level taxonomy of 197 localizable repairs, comprising 10 repair groups and 60 action patterns, and characterize the complete set of 221 fix-bearing issues. The analysis identifies concentrations in \emph{Startup and Environment}, \emph{Tool Execution}, \emph{Tool Schema and Discovery}, and \emph{Documentation}, as well as differences in patch structure across repair groups.

\item \textbf{MCP-RepairBench.} We construct MCP-RepairBench, comprising 60 real repair tasks from 25 repositories. Each task combines pinned buggy and developer-fixed revisions with a case-specific behavioral evaluator and reproducible offline environment, with preservation checks where applicable to constrain deletion, hard-coding, and incomplete one-surface repairs. The benchmark is designed to determine whether a repair restores the behavior required by the issue rather than merely satisfying a limited test suite.

\item \textbf{Evaluation of automated MCP repair.} We evaluate an MCP-aware repair agent with evaluator feedback under a predeclared experimental configuration and report its effectiveness, API cost, resource use, and recurring failure mechanisms. The agent resolves 34/60 tasks on its first attempt and 41/60 within three independently initialized attempts. Execution traces identify configuration and package repair, exact response and lifecycle requirements, response-limit constraints, and probe selection as continuing challenges.

\end{itemize}

\noindent\textbf{Replication packages.}
The data, codebooks, metrics, analysis scripts, and generated results for RQ1 and RQ2 are available in the empirical-study replication package~\cite{mcpRepairStudyArtifact}. MCP-RepairBench is available as release \texttt{v1.0.0}~\cite{mcpRepairBenchArtifact}, while the evaluated repair agent, baseline, experimental configurations, outcomes, patches, traces, and analysis scripts are available as repair-agent package release \texttt{v1.0.1}~\cite{mcpRepairAgentArtifact}.

\noindent The remainder of the paper presents the background (Section \ref{sec:background}), methodology (Section \ref{sec:method}), and results (Section \ref{sec:result}), followed by a discussion of implications (Section \ref{sec:discuss}), related work (Section \ref{sec:related_work}), threats to validity (Section \ref{sec:threat}), and future research (Section \ref{sec:conclusion}).

\section{Background}
\label{sec:background}
This section reviews MCP architecture, bug-fix pattern abstractions, and repository-level repair and behavioral validation.

\subsection{MCP Architecture and Integration Boundaries}
The Model Context Protocol (MCP) standardizes how model-based applications exchange context and invoke external capabilities. An MCP host creates a client for each connected server; the client and server establish a stateful session, negotiate capabilities, and exchange JSON-RPC messages. Servers can expose tools, resources, and prompts, while clients can provide features such as sampling and filesystem roots. MCP separates this data layer from the transport layer, which carries messages over local standard input/output (stdio) or Streamable HTTP~\cite{mcpSpec2025}. A server implementation must therefore coordinate protocol state with its SDK, host behavior, invoked services, authentication and configuration, packaging, and the runtime environment.

These dependencies make MCP servers software components whose correctness depends not only on their local implementation but also on their integration with clients, protocols, tools, and deployment environments. In this sense, MCP servers can be viewed as \emph{integration-boundary software}: their behavior is determined by contracts and interactions that extend beyond the server itself. Consequently, code that is locally valid can still fail when integrated with an MCP client or deployed in a target environment. For example, a server may advertise a tool schema that is inconsistent with the function handling tool requests, return content in a format that the client cannot decode, write logs or error messages to \texttt{stdout} and thereby interfere with MCP messages, or rely on configuration that is unavailable in the target deployment environment.

\subsection{Bug-Fix Pattern Abstractions}
A bug-fix pattern abstracts a recurring repair operation from individual patches. Existing studies derive patterns from syntactic edits and their context, nested change structures, or large collections of fine-grained repair actions~\cite{panFixPatterns2009,islamFixPatterns2020,camposFixPatterns2019}. These abstractions are useful for characterizing how developers repair faults and for constructing repair operators. However, repair vocabularies can vary across programming languages and application domains. For example, studies of Python and deep learning (DL) systems have identified repair patterns and challenges that are not adequately captured by conventional Java statement-level fix-pattern taxonomies~\cite{yangPythonFixPatterns2022,islamDNNRepair2020}.
Pattern abstractions may be defined at different levels, from recurring syntactic edits to semantic descriptions of a repair's operation and intent. At the semantic level, syntactically different patches may represent the same repair when they establish the same externally observable behavior for the same corrective purpose. Conversely, superficially similar edits may represent different repairs when they serve different purposes. This semantic view enables repair patterns to be analyzed across implementation idioms, programming languages, and artifact types without equating a descriptive pattern with a patch-generation template.

\subsection{Repository-Level Repair and Behavioral Validation}
Automated program repair (APR) generates candidate modifications from evidence such as failing tests, specifications, or failure reports~\cite{monperrus2018,gazzola2019}. Code language model studies extend this space by generating patches directly, while also highlighting the need for realistic execution-based evaluation and controls against training-data leakage~\cite{jiangCodeModelsAPR2023}. Repository-level issue resolution broadens the repair problem further: a repair system must interpret natural-language evidence, locate the relevant artifacts, modify source and non-source files, and validate the resulting repository state~\cite{swebench2024}. Contemporary repair workflows therefore organize these activities into an inspect--diagnose--edit--validate loop, often using execution feedback to iteratively refine candidate repairs~\cite{chatrepair2024,sweagent2024,repairagent2024}.

The scope of validation ultimately constrains what such evaluations can establish. A patch that passes the available tests is \emph{plausible}, but not necessarily correct, because an incomplete oracle may accept overfitting or behaviorally divergent repairs~\cite{qi2015,swebenchCorrectness2026}. At the repository level, validation may therefore need to establish not only that the reported failure is addressed, but also that expected behavior is preserved across interacting artifacts and execution boundaries, including source code, processes, transports, configuration, packaging, and external services.

\section{Methodology}
\label{sec:method}
\begin{figure*}[t]
\centering
\includegraphics[width=\textwidth]{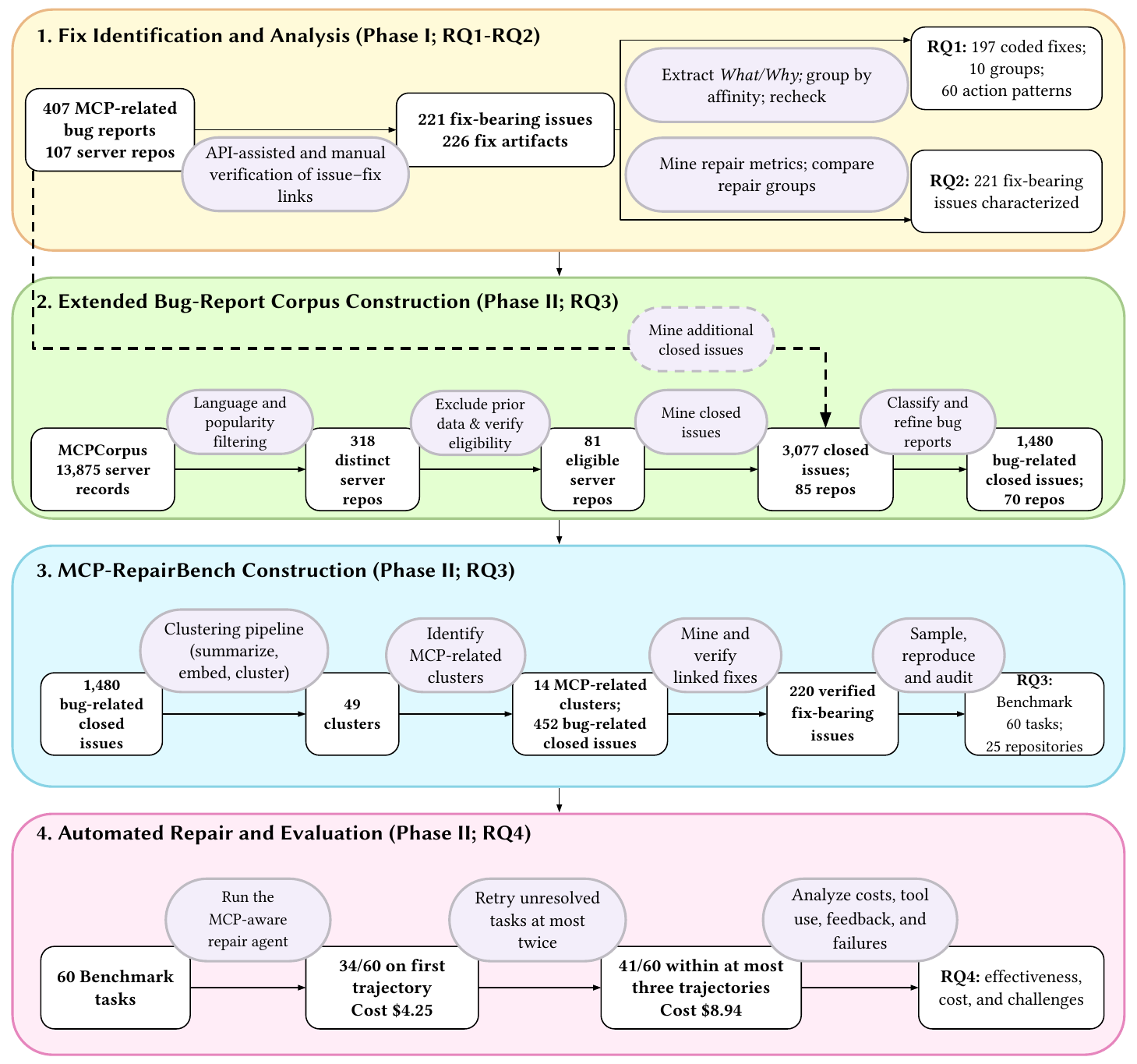}
\caption{Study overview. Phase~I identifies verified repairs in the original MCP fault dataset and analyzes their semantic patterns and characteristics (RQ1--RQ2). Phase~II extends the evidence base with closed issues from newly eligible MCPCorpus repositories and newly available closed issues from the 107 Phase~I repositories; the dashed path denotes the latter contribution. The resulting corpus supports MCP-RepairBench construction and audit (RQ3), followed by evaluation and trace analysis of the MCP-aware repair agent (RQ4). In RQ4, each independently initialized repair attempt is a trajectory.}
\Description{Four horizontal methodology stages. Stage 1 starts from 407 bug reports in 107 server repositories, identifies 221 fix-bearing issues represented by 226 fix artifacts, and branches into RQ1 semantic repair coding and RQ2 metric analysis. Stage 2 filters 13,875 MCPCorpus server records to 81 newly eligible servers and combines their closed issues with newly available closed issues from the 107 Phase~I repositories, shown by a dashed path. The combined collection contains 3,077 closed issues from 85 contributing repositories and is refined to 1,480 bug-related closed issues from 70 repositories. Stage 3 forms 49 clusters, identifies 14 MCP-related clusters containing 452 closed issues, verifies linked fixes for 220 closed issues, and produces 60 reviewed MCP-RepairBench tasks from 25 repositories for RQ3. Stage 4 evaluates the MCP-aware repair agent, reports pass@1 of 34/60 and pass@3 of 41/60, and analyzes cost, tool use, feedback, and failures for RQ4.}
\label{fig:study_overview}
\end{figure*}

This section presents the study as a staged empirical investigation comprising two phases. Phase~I identifies and characterizes developer repairs using a previously validated MCP fault dataset (RQ1--RQ2). Phase~II uses a separate extended bug-report corpus to construct MCP-RepairBench from real fix-linked cases (RQ3) and evaluates an MCP-aware repair agent on the resulting benchmark (RQ4). Figure~\ref{fig:study_overview} distinguishes the units of analysis and roles of the two phases. The findings from Phase~I also inform Phase~II: the identified repair groups and action patterns guide coverage-oriented benchmark selection, while the observed integration boundaries inform the design of MCP-specific tools for the repair agent.

\subsection{Primary dataset and fix identification}\label{fix_identification}
The starting dataset is the replication artifact released by Taraghi et al.~\cite{ourPaper}, containing 407 MCP-related GitHub closed issues extracted from 107 MCP servers. In their fault-taxonomy analysis, they assigned 419 fault labels to 383 of these issues, including 36 issues with two labels; the remaining 24 were not assigned a fault label because the authenticity of the fault or its root cause could not be established. Because the artifact does not identify how the reported problems were repaired, we examine all 407 issues for fixing pull requests (PRs) and commits. We use two operational definitions in this study. A \emph{fix artifact} is a PR or direct commit that implements all or part of an issue's repair. A \emph{fix-bearing issue} is an issue for which we verify at least one such artifact. Our identification procedure has two stages. API-assisted candidate discovery and validation is followed by an exhaustive manual search of issues for which the API query yields no candidate.

Prior bug-dataset construction studies motivate an API-assisted first stage~\cite{songBugMiner2023,niMegaVul2024}. 
We use GitHub's REST API~\cite{github_api_v3} to extract detailed information regarding each issue, specifically the issue timeline. 
We inspect timeline records whose \texttt{event} field is \texttt{`cross-referenced'}, \texttt{`closed'}, \texttt{`referenced'}, or \texttt{`connected'}, because these event types can expose associated commits or PRs~\cite{githubIssueEventTypes2026}. The query yields at least one candidate artifact for 102 issues.  
Although earlier research on mining software repositories linked issue reports to code changes through explicit issue identifiers in change logs and repository metadata~\cite{fischerReleaseHistory2003}, subsequent studies showed that such explicit links can be incomplete or systematically biased and recovered additional candidates using temporal proximity, textual similarity, and developer information~\cite{bachmannMissingLinks2010,wuReLink2011}.
To this end, we treat the gathered associations in this step as candidates rather than verified fixes.

Accordingly, we validate each candidate against the issue discussion and its development context, including the diff and the PR title and description or commit message. We retain an association only when the discussion or artifact metadata connects it to the report and inspection of the changed files confirms that the patch addresses the reported problem. Applying these criteria verifies fixes for 99 issues in the first stage. In two cases, a cross-referenced PR does not implement the reported fix, but the issue discussion identifies the implementing commit; we retain that commit instead of the unrelated PR.

In the second stage, we manually search for repairs that were not surfaced by the API query. The first two authors divide the remaining 305 issues for exhaustive inspection of their discussions and repository histories. They search for four forms of repair evidence. These are (i) a PR identified as the fix, (ii) a linked or named fixing commit, (iii) a version identified as containing the fix, or (iv) a maintainer statement that the repair was pushed in a recent commit or release. Directly linked PRs and commits are checked against their diffs. For version- and release-based evidence, the authors inspect the relevant tag and temporally adjacent commits, retaining an artifact only when its message, issue reference, release history, or code change establishes both a semantic and temporal match to the reported problem.
For ambiguous associations, the first two authors jointly review each case in meetings and retain only those for which they reach agreement on a verified issue--fix link.
The second-stage manual search identifies another 122 fix-bearing issues, yielding 221 across both stages.

The final dataset contains 221 fix-bearing issues represented by 226 fix artifacts (Table~\ref{tab:fixes}). A fix-bearing issue is the unit of RQ1--RQ2 analysis. When a candidate commit belongs to a PR, we record the PR because it preserves the complete patch and its review and merge context. We retain multiple artifacts for an issue only when each implements a distinct necessary part of the repair; five issues have two such artifacts, accounting for the difference between the issue and artifact totals. For example, two PRs jointly repair \textit{awslabs/mcp} issue~\#82, with one correcting the referenced server and the other removing a trailing comma from the configuration example.\footnote{\url{https://github.com/awslabs/mcp/issues/82}, \url{https://github.com/awslabs/mcp/pull/80}, and \url{https://github.com/awslabs/mcp/pull/83}.}

\begin{table}[htbp]
\caption{Verified fix evidence in the original MCP fault dataset.}
\label{tab:fixes}
\centering
\small
\begin{tabular}{lrr}
\toprule
\textbf{Fix representation} & \textbf{Fix-bearing issues} & \textbf{Fix artifacts} \\ \midrule
Pull request & 148 & 152 \\
Direct commit & 73 & 74 \\ \midrule
\textbf{Total} & \textbf{221} & \textbf{226} \\ \bottomrule
\end{tabular}
\end{table}

\subsection{Open coding and semantic repair synthesis}\label{subsec:semantic_repair_coding}
To answer RQ1, we derive a two-level repair taxonomy in two distinct qualitative stages. We first conduct independent open coding of repair evidence and then use collaborative affinity diagramming.

Existing fix-pattern taxonomies primarily describe recurring source-code edits and do not consistently cover repairs to configuration, dependencies, build files, documentation, and MCP interfaces~\cite{panFixPatterns2009,islamFixPatterns2020,camposFixPatterns2019}. We therefore use open coding so that the repair labels emerge through systematic comparison of the issue reports and their corresponding patches. This design follows methodological guidance for empirical software-engineering research~\cite{seaman1999,saldanaCoding2021} and has precedent in the open coding of domain-specific repair patterns~\cite{islamDNNRepair2020}. We characterize the resulting repairs by their semantic operation and intent rather than by language-specific edit templates.

The unit of analysis in this step is one verified issue--repair record. During open coding, the first two authors independently inspect the issue report, discussion, changed files, and diff for each record and answer two analytical questions. \emph{1) What changed in the fix?} captures the observable repair operation performed by the patch, whereas \emph{2) Why was it changed?} captures the failure mechanism or corrective rationale supported by the issue and patch evidence. We refer to the resulting open codes as the \emph{What} and \emph{Why} descriptions. This separation prevents syntactically different implementations of the same repair from being divided solely by edit form, while also preventing superficially similar edits with different purposes from being conflated. The questions are an evidence-extraction structure, not a claim that \emph{What} and \emph{Why} constitute an established bug-fix-pattern taxonomy; they provide a bridge from concrete changes to semantic repair operations, consistent with software-change research that distinguishes modifications from their rationale and underlying intention~\cite{alsafwanRationale2022,kruegerChangeIntentions2024}.

We examine all 221 issue--repair records in 11 batches. Ten batches contain 20 records, and the final batch contains 21. For each batch, the first two authors independently inspect every record and attempt to formulate its \emph{What} and \emph{Why} descriptions. At the end of each batch, they meet to compare their descriptions and resolve disagreements by returning to the underlying evidence. They also revise the wording and boundaries of provisional codes when necessary and merge codes that describe the same repair operation or rationale before proceeding to
the next batch.
This iterative process follows batch-wise taxonomy-construction procedures used in prior empirical software-engineering studies~\cite{humbatovaFaultTaxonomy2020,morovatiBugCharacterization2024}. After all batches have been examined, the authors reconcile the complete coding table and consolidate the description-level codebook. If the two coders cannot reach consensus on a coding decision, the third author adjudicates the case, consistent with the protocol reported by similar studies~\cite{shah2026characterizing,morovati2024common}.

A record is retained in the open-coded dataset only when its evidence supports one sufficiently localizable repair interpretation. Applying this criterion retains 197 of the 221 fix-bearing issues. The other 24 remain verified fixes but are not assigned \emph{What} and \emph{Why} descriptions because their patches are too large, diffuse, or multi-purpose to isolate one specific repair operation confidently. 
Their median patch changes 1,732.5 Lines of Code (LOC) across 16.5 files, compared with 46 LOC and two files among the coded repairs. For example,
the repair associated with \textit{NiFiMCP} issue~\#14 changes 13,134 LOC across
66 files and includes changes to configuration, documentation, user
interfaces, server and workflow code, and tests; we could not isolate
one issue-specific repair operation from this broader artifact with
sufficient confidence.\footnote{Issue and fix:
\url{https://github.com/ms82119/NiFiMCP/issues/14} and
\url{https://github.com/ms82119/NiFiMCP/commit/b8a230052c59177a9ea0d28e4315bfc0751c4750}.}
These 24 records remain in the overall RQ2 summaries but are excluded from the taxonomy and repair-group comparisons.

After completing open coding and reconciling the 197 retained \emph{What} and \emph{Why} descriptions, the authors use affinity diagramming to synthesize them into the taxonomy. Affinity diagramming externalizes qualitative observations and organizes them into emergent groups through repeated comparison and rearrangement~\cite{luceroAffinity2015}; empirical software-engineering studies have used it to synthesize code-review evidence, software-architecture risks, and inductively generated codes into higher-level categories~\cite{bacchelliBirdCodeReview2013,bassArchitectureAffinity2008,foundjemOnboarding2021}. In our application, semantically similar repair operations and rationales first form action patterns, and related action patterns subsequently form broader repair groups. During synthesis, records and provisional groups may be moved, merged, or split; ambiguous cases are resolved by returning to the issue and patch evidence. The authors finally name and define both taxonomy levels, document their inclusion boundaries, and recheck every retained record against the final codebook. This iterative consolidation follows established guidance for developing qualitative coding schemes~\cite{seaman1999,campbellCoding2013}.

Because no predefined labels exist and the codes and their boundaries evolve through batch reconciliation, an agreement coefficient calculated during the process would not measure independent application of a stable codebook~\cite{saldanaCoding2021}.
Thus, we do not report an intercoder-agreement coefficient for this formative open-coding stage. 
We therefore report the independent-coding, reconciliation, adjudication, and final-definition recheck procedures explicitly; independent validation using the completed fixed codebook remains a limitation discussed in Section~\ref{sec:threats}. The resulting taxonomy contains 10 repair groups and 60 action patterns; complete definitions and counts are reported with RQ1 and in the appendix.

\subsection{Repair metrics and statistical analysis}\label{subsec:repair_metrics}
We organize the RQ2 measures into three complementary groups established in prior empirical software-engineering studies. These are structural fix metrics, process metrics, and developer-context metrics~\cite{gigerFixTime2010,rahmanExperience2011,gousiosPullBased2015,morovatiBugCharacterization2024}. The structural metrics characterize the magnitude and dispersion of the patch; the process metrics capture elapsed time in reporting repair and integration; and the developer-context metrics approximate the fixer's prior contribution experience. It should be taken into account that no single metric is treated as a direct estimate of repair effort. For each fix-bearing issue, we obtain the required timestamps, authors, commits, changed files, additions, and deletions through GitHub's REST API~\cite{github_api_v3}.

\paragraph{Structural fix metrics.} 
We characterize each fix using changed LOC, changed files, commit count, and change entropy, following established change-level measures in software-engineering studies~\cite{ekenChangeMetrics2021,morovatiBugCharacterization2024}. Changed LOC is calculated as the sum of added and deleted lines; changed files and commit count record the number of modified files and commits, respectively.
\emph{Change entropy} measures how evenly the changed lines are distributed across the modified files. This metric is based on Shannon's information entropy~\cite{shannonInformation1948} and follows its application to code-change complexity by Hassan and subsequent software-engineering studies~\cite{hassanChangeEntropy2009,chaturvediChangeEntropy2014,ekenChangeMetrics2021,morovatiBugCharacterization2024}. Let $c_i$ denote the changed-line count for file $i$, and let $p_i=c_i/\sum_j c_j$ denote that file's share of all changed lines in the repair. We compute $H=-\sum_i p_i\log_2p_i$. A value of zero indicates that all changed lines are concentrated in one file, while higher values indicate that the change is distributed more evenly across multiple files. If a repair contains no changed lines, we assign its entropy a value of zero. All four structural metrics are available for all 221 fixes.

\paragraph{Process metrics.} Established fix-time and PR studies measure \emph{completion latency} as the interval from issue creation to PR merge or direct-commit time, and \emph{PR duration} as the interval from PR creation to merge~\cite{gigerFixTime2010,gousiosPullBased2015}. 
In 11 MCP-related bugs, the corresponding issue reports were created after the selected fixes had already been completed.
To preserve the integrity of the dataset, we retain these cases in the structural summaries; however, their negative completion latencies are excluded, leaving 210 fixes in the completion-latency analysis. It is also worth noting that PR duration is computed only for the 148 issues associated with an applicable merged PR. The remaining issue reports are excluded from this analysis rather than being assigned an artificial duration of zero, thereby avoiding bias in the reported results.

\paragraph{Developer-context metrics.} Prior-commit counts are established proxies for developer experience in previous similar studies~\cite{rahmanExperience2011,ekenChangeMetrics2021,morovatiBugCharacterization2024}. Accordingly, we measure \emph{repository experience} as the fixer's number of prior commits in the focal repository before the fix endpoint and \emph{global experience} as the fixer's number of prior GitHub commits in any repository before that endpoint. The 25 bot-authored fixes remain in the structural and process summaries but are excluded from experience analyses, leaving experience values for 196 fixes.

\paragraph{Multiple fix artifacts.} For an issue represented by more than one fix artifact, we aggregate all verified artifacts into one issue-level observation. We sum commit counts and changed LOC, combine repeated changes to the same file, count each modified file once, and recompute entropy from the resulting file-level distribution. For multi-artifact repairs, completion latency ends at the latest required artifact endpoint; when the artifacts are merged PRs, PR duration spans from the creation of the earliest PR to the merge of the latest, while commit-only repairs have no PR duration. When a repair has multiple human fixers, we count each author once and average their repository and global experience values.

\paragraph{Statistical analysis.} 

Descriptive summaries use all 221 fix-bearing issues, whereas repair-group comparisons use the 197 coded repairs, subject to the metric-specific exclusions described above. Within the coded subset, all 197 repairs contribute to the structural comparisons. The same exclusion rules leave 188 repairs for completion latency, 133 for merged-PR duration, and 173 for each fixer-experience measure. We do not compare the 60 action patterns statistically because 19 contain only one repair and another 11 contain only two. Thus, half of the patterns have at most two observations. Our statistical analysis examines correlations across the structural, process, and developer-context metrics and compares the eight repair metrics across repair groups. We use a significance level of $\alpha=0.05$.

For every metric, the \textit{Shapiro--Wilk} test rejects normality in each applicable analysis sample ($p<0.001$)~\cite{shapiroWilk1965}.

To examine whether the three dimensions of fix characteristics are correlated, we compute two-sided \textit{Spearman} rank correlations between metrics from different families~\cite{spearman1904}. The analysis includes eight structural--process, eight structural--experience, and four process--experience comparisons, yielding 20 tests. %
We do not include correlations between metrics within the same family because the analysis focuses on relationships across structural, process, and developer-context characteristics. Rank correlation is appropriate because the metrics are non-normally distributed and contain tied observations. Each correlation uses the fixes for which both metrics are available. Comparisons of structural metrics with completion latency therefore use 210 fixes, and those with merged-PR duration use 148. Comparisons of structural metrics with fixer experience use 196 fixes, while process--experience comparisons use 186 fixes for completion latency and 131 for merged-PR duration. We apply the Bonferroni correction across all 20 tests.

Given the non-normal distributions, we use the \textit{Kruskal--Wallis} test to determine whether a metric differs across the repair groups~\cite{kruskalWallis1952}. When this test identifies a statistically significant difference, we conduct post-hoc pairwise comparisons using \textit{Dunn's test}~\cite{dunn1964} with Bonferroni correction~\cite{dunn1961}. We report $\eta^2=H/(N-1)$ for each Kruskal--Wallis comparison~\cite{morseEffectSize1999} and directional Vargha--Delaney $A_{12}$ for significant pairwise comparisons~\cite{varghaDelaney2000}. We interpret $\eta^2$ values of at least 0.01, 0.06, and 0.14 as small, medium, and large, respectively~\cite{morseEffectSize1999}. For $A_{12}$, values of 0.56--0.64 (or 0.36--0.44) indicate a small effect, 0.64--0.71 (or 0.29--0.36) a medium effect, and values above 0.71 (or below 0.29) a large effect~\cite{varghaDelaney2000}.

\subsection{Extended bug corpus and fix-linked candidate discovery}
Phase~II begins by extending the closed-issue corpus and identifying MCP-related closed issues with verified repairs. The subsequent sections describe how these fix-linked candidates become reviewed benchmark tasks and how the repair agent is evaluated on the resulting benchmark.
It is worth noting that the issue--fix pairs identified in Phase~I are not reused as evaluation tasks.
Indeed, reusing these issue--fix pairs
would evaluate the agent on evidence that had already informed its general prompt and tool-family design, weakening the separation between development and evaluation. 
A distinct issue pool also broadens the temporal and repository coverage of the benchmark beyond the original corpus.

To keep the two datasets comparable, Phase~II retains four elements of the data-collection protocol reported by Taraghi et al.~\cite{ourPaper}: the repository eligibility rules, the definition and collection of closed issues, the issue-classification prompt and model configuration, and the semantic-clustering workflow and configuration, including issue summarization, embedding, dimensionality reduction, cluster-structure estimation, and final cluster assignment. The applied thresholds, collection windows, models, and key algorithm configurations are reported at the corresponding stages below.

MCPCorpus supplies the external source of new candidates~\cite{mcpCorpus2025}, 
providing 13,875 MCP server records with associated GitHub metadata as of June~3, 2025.
Repository discovery combines automated metadata filtering with manual eligibility review. To preserve language and SDK consistency with Phase~I, we retain records whose reported primary language is Python or Jupyter Notebook and require at least 10 stars and 10 forks.
We convert each GitHub URL to its \texttt{owner/repository} identifier and remove duplicate records that point to the same repository. After this step, the language and popularity filters yield 318 distinct repositories.
To avoid overlap with the earlier study, we compare the Phase~II candidates with all 130 server repositories retained in that study's corpus. Removing these 130 overlapping repositories leaves 188 previously unstudied repositories for Phase~II. Of the 130 repositories in the earlier corpus, 107 contributed the issues analyzed in Phase~I.
Each remaining repository must have at least one closed issue and one merged PR, providing evidence of both issue-reporting activity and accepted code changes. This screen retains 112 repositories. The \emph{Repository discovery} stage of Table~\ref{tab:benchmark_funnel} reports every intermediate filtering step from MCPCorpus records to repository candidates.

The first two authors then manually inspect the 112 candidate repositories to verify whether they implement genuine Python MCP servers. Specifically, we exclude repositories that use MCP solely as a client, proxy, framework, or in tests
as well as repositories that consist only of tutorials, templates, or sample projects. This manual verification step excludes 31 repositories, resulting in a final set of 81 eligible MCP server repositories for subsequent analysis.

Issue collection for this project concluded on May~11, 2026, which serves as the common upper bound for both repository sources. For the 81 newly eligible repositories, we retrieve closed, non-PR issues created from November~1, 2024 through the collection cutoff. This lower bound covers the period surrounding Anthropic's public introduction of MCP on November~25, 2024~\cite{anthropicMCPLaunch2024}. For the 107 repositories of Phase~I, we retrieve closed issues created after the latest issue-creation date represented in the Phase~I artifact, thereby extending rather than duplicating its temporal coverage. We retain only URLs absent from that artifact and deduplicate the combined collection by URL. The result comprises 3,077 closed issues from 85 contributing repositories.

\FloatBarrier
\begin{table}[htbp]
\small
\caption{Detailed accounting of the number and type of items remaining after each benchmark filtering and construction step.}
\label{tab:benchmark_funnel}
\centering
\setlength{\tabcolsep}{4pt}
\begin{tabularx}{\columnwidth}{>{\centering\arraybackslash}p{0.21\columnwidth}>{\raggedright\arraybackslash}Xr}
\toprule
\textbf{Stage} & \textbf{Filtering or construction step} & \textbf{\# items} \\ \midrule
\multirow{7}{*}{\parbox{0.19\columnwidth}{\centering\emph{Repository discovery}}}
& MCPCorpus server records & 13,875 \\
& Python or Jupyter Notebook server records & 4,568 \\
& Server records with at least 10 stars and 10 forks & 399 \\
& Distinct GitHub repositories after deduplication & 318 \\
& New repositories after excluding original-dataset repositories & 188 \\
& Repositories satisfying closed-issue and merged-PR activity criteria & 112 \\
& New MCP server repositories after manual eligibility review & 81 \\ \midrule
\multirow{5}{*}{\parbox{0.19\columnwidth}{\centering\emph{Issue discovery and candidate identification}}}
& Closed non-PR issues after collection and URL deduplication & 3,077 \\
& Closed issues classified as bug related & 1,620 \\
& Bug-related closed issues after refinement & 1,480 \\
& Closed issues in 14 MCP-related clusters & 452 \\
& Closed issues with verified fixes after automatic and manual discovery & 220 \\ \midrule
\multirow{4}{*}{\parbox{0.19\columnwidth}{\centering\emph{Benchmark selection and construction}}}
& Fix-linked closed-issue candidates in the initial manual selection & 60 \\
& Initial benchmark candidates retained after case-level audit & 44 \\
& Replacement benchmark candidates retained after the same reviews & 16 \\
& \textbf{Released benchmark tasks from 25 repositories} & \textbf{60} \\ \bottomrule
\end{tabularx}
\end{table}

To preserve label consistency across the two datasets, we reuse the issue-classification prompt and model configuration reported by Taraghi et al.~\cite{ourPaper}, including the selected model snapshot, \texttt{gpt-4o-mini-2024-07-18}~\cite{openai_gpt4omini_launch}. Only closed issues classified as bugs proceed to the subsequent quality screen. The classifier identifies 1,620 bug-related closed issues. Because automated classification can be sensitive to project labels and issue report wording~\cite{kallisIssueTypes2021}, we subsequently apply a rule-based validation using issue metadata and text obtained through GitHub's REST API~\cite{github_api_v3}. 
This validation step flags reports whose \texttt{state\_reason} is \texttt{not\_planned} and reports for which the phrase ``\textit{duplicate of}'' appears in the title, body, or discussion; GitHub labels are also consulted during the subsequent manual review.
These rules flag 148 reports, including 133 closed as not planned and 15 containing duplicate text. We manually inspect all flagged reports and restore eight false positives for which the discussion, linked repair, labels, or repository state establishes a genuine fault. The refinement therefore excludes 140 reports and retains 1,480 unique bug-related closed issues from 70 repositories. These collection, classification, and refinement steps correspond to the \emph{Issue discovery and candidate identification} stage of Table~\ref{tab:benchmark_funnel}.
The replication package records the queries, issue URLs, classifier configuration, flags, and review decisions to preserve the chain of evidence~\cite{mcpRepairBenchArtifact}.

We organize the 1,480 bug-related closed issues using the inherited 
semantic clustering workflow. The purpose of this step here is to structure manual review and candidate discovery, not to construct another fault or repair taxonomy. The workflow uses \texttt{gpt-4.1-mini-2025-04-14} to summarize the problem, context, and resolution evidence in each issue's title, body, and comments~\cite{openai_gpt41mini_model,ramirezIssueComments2020}. 
We encode these summaries with \texttt{gemini-embedding-001}~\cite{geminiEmbeddingDocs}. We then apply UMAP for dimensionality reduction~\cite{umap2018} and HDBSCAN to estimate the grouping structure without specifying the number of clusters in advance~\cite{hdbscan2013}. The resulting value of $k=49$ is supplied to KMeans so that every issue receives a cluster assignment~\cite{macqueen1967}. This sequence adopts the modular design of BERTopic and established reporting guidance for topic-based analysis of software-engineering text~\cite{grootendorstBERTopic2022,costaSilvaTopicModeling2021}.

The authors review representative summaries, cluster keywords, and sampled reports to identify clusters involving MCP server behavior. Fourteen clusters, containing 452 unique closed issues, meet this criterion and form the candidate pool for fix-link mining.

We apply the fix-identification procedure described in Section~\ref{fix_identification} to this new dataset of 452 closed issues. GitHub timeline events automatically identify candidate fix links for 86 of the 452 closed issues~\cite{githubTimelineEvents2026}; manual verification confirms 85 of them. We then inspect the remaining issues for PRs or commits referenced in their discussions or repository histories. This manual search identifies another 150 candidate fixes, of which 135 pass verification. In total, we retain verified fixes for 220 of the 452 closed issues (48.7\%).

\subsection{MCP-RepairBench selection, construction, and quality review}
The 220 closed issues with verified fixes form the pool from which MCP-RepairBench is selected and constructed. 
This subsection describes the initial selection, case construction, two-stage quality review, and final release verification.

\subsubsection{Initial candidate selection}
Fix identification ultimately yields 220 verified issue--fix pairs. We use purposive maximum-variation selection to construct the initial benchmark candidate set. This non-random method supports the deliberate selection of information-rich cases that differ along dimensions relevant to the study objective, while allowing researchers to apply domain knowledge during selection~\cite{patton2015,baltesSampling2022}. It is appropriate here because MCP-RepairBench is intended to exercise varied repair behaviors rather than reproduce their prevalence in the wider ecosystem. Prior software-engineering research has similarly emphasized diversity when selecting study objects and has used purposive selection when exhaustive evaluation is impractical~\cite{nagappanDiversity2013,laaberMicrobenchmarks2021}. The first two authors review a coverage-oriented ranking of the available candidates. Each candidate is characterized by its repository, affected MCP behavior, repair operation, anticipated method of evaluation, issue clarity and reproducibility, strength of the issue–fix evidence, MCP relevance, and potential construction risks, such as reliance on external services or platform-specific behavior. Selection first retains strong candidates spanning the available MCP behaviors and repair operations. It then fills a 52-case core by favoring candidates with stronger evidence that also improve coverage of underrepresented behaviors, repair operations, or repositories, while limiting repeated selection from the same repositories. Finally, eight candidates are added to strengthen coverage of external-service integration, connection and transport, and platform and build behavior. This procedure produces the initial 60 candidates submitted to the construction and review process described below.
For reproducibility, the replication package records the issue and fix links and coverage rationale for the initial selection, together with the subsequent construction and review records~\cite{mcpRepairBenchArtifact}.
The transition from the 220 closed issues with verified fixes to the audited 60-task release is reported in the \emph{Benchmark selection and construction} stage of Table~\ref{tab:benchmark_funnel}.

\begin{figure}[htbp]
\centering
\includegraphics[width=\textwidth]{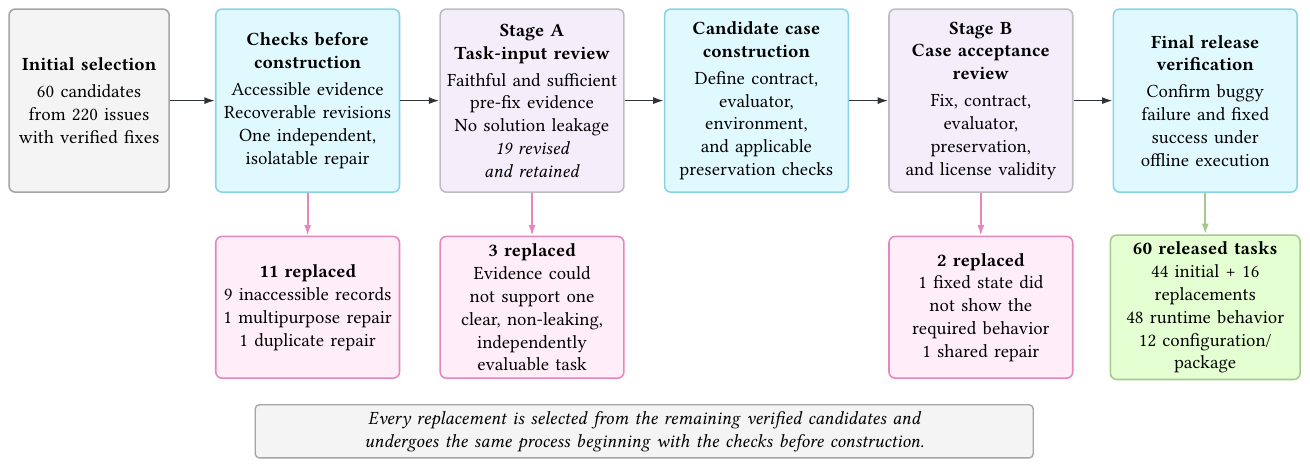}
\caption{Construction and review of MCP-RepairBench. The pink boxes report replacements at the step where the corresponding candidates were excluded. Revised task descriptions that remained in the benchmark are not counted as replacements. Every replacement undergoes the same checks and reviews as an initially selected candidate.}
\Description{A left-to-right pipeline begins with 60 initially selected candidates and proceeds through checks before construction, Stage A task-input review, candidate case construction, Stage B case acceptance review, and final release verification. Pink boxes show 11 replacements before construction, three during Stage A, and two during Stage B. The process produces 60 released tasks, including 44 from the initial selection and 16 replacements.}
\label{fig:benchmark_construction_pipeline}
\end{figure}

The initial set of 60 candidate issues does not by itself determine which cases are ultimately included in the released benchmark. 
As Figure~\ref{fig:benchmark_construction_pipeline} shows, each candidate passes through checks before construction, task-input review in Stage~A, candidate case construction, case acceptance review in Stage~B, and final release verification. When a candidate is excluded, the first two authors select a replacement from the remaining issues with verified fixes. Every replacement follows the same process from the beginning.

\subsubsection{Initial checks before construction}
The initial selection is based on the mined records and evidence available during candidate screening. Before preparing a task input, we confirm that the public issue and linked repair remain accessible, that the buggy and fixed revisions can be recovered, and that the repair can be represented as one independent case. These checks replace 11 candidates. Nine belong to a repository whose issue and repair pages became inaccessible after data collection.\footnote{\url{https://github.com/doobidoo/mcp-memory-service}.} One uses a multipurpose repair that cannot be isolated as one case.\footnote{Issue and repair: \url{https://github.com/doobidoo/mcp-memory-service/issues/173} and \url{https://github.com/doobidoo/mcp-memory-service/commit/0008b32f4a0106ab9c205dda21be76658c8f731b}.} The remaining candidate duplicates another repair.\footnote{The repair for \url{https://github.com/ClickHouse/mcp-clickhouse/issues/71} is already represented by \url{https://github.com/ClickHouse/mcp-clickhouse/issues/143}; both refer to \url{https://github.com/ClickHouse/mcp-clickhouse/pull/145}.} 

\subsubsection{Stage A: task-input review}
For every candidate that passes these checks, we prepare an \emph{issue-derived task description} using evidence available before the repair was developed. Its content depends on the evidence provided for that issue. We retain the information needed to understand and reproduce the reported failure, which may include the issue title and body, reproduction steps, observed and expected behavior, logs or stack traces, relevant versions and configuration, attachment transcriptions, and pre-fix comments that clarify the failure. We exclude post-fix discussion and any confirmed root cause, proposed solution, exact implementation location, or PR/commit link that would reveal the developer repair.

One author reviews only this proposed description and its eligible pre-fix evidence. The review asks whether the description faithfully represents the issue, provides enough information to attempt a repair, and avoids solution leakage. These checks address problems observed in prior repair benchmarks, where some issue descriptions provide insufficient information while others disclose exact or misleading solution steps~\cite{agentless2024}. The review form records the decision, whether more pre-fix discussion is needed, and explanatory notes. Nineteen descriptions are revised to remove solution information and pass a second review. Three candidates cannot support a sufficiently clear and non-leaking task based on one independent repair and are replaced. For example, issue~\#65 in \textit{ROS MCP Server} does not provide enough information on its own, while the additional discussion required to clarify the failure also discloses the troubleshooting solution.\footnote{\url{https://github.com/robotmcp/ros-mcp-server/issues/65}}
Stage~A ends when the task description is approved and locked. Figure~\ref{fig:benchmark_task_examples} shows two resulting task descriptions. They illustrate how the retained evidence can range from a short error report to a report supplemented with an attachment transcription.

\begin{figure}[htbp]
\centering
\begin{tcbraster}[raster columns=2,raster equal height=rows,raster column skip=6pt]
\begin{tcolorbox}[enhanced,colback=white,colframe=benchmarkexampleframe,colbacktitle=benchmarkruntime,
  coltitle=black,fonttitle=\bfseries,title={(a) Case 0213: runtime behavior},
  boxrule=0.5pt,arc=2pt,left=5pt,right=5pt,top=4pt,bottom=4pt,before upper=\raggedright]
\small
\textbf{Issue:} \href{https://github.com/semgrep/mcp/issues/154}{semgrep/mcp\#154}

\medskip
\textbf{Title}\par
When running \texttt{semgrep\_scan\_local} Error running semgrep scan: \texttt{NoneType} object has no attribute \texttt{decode}

\medskip
\textbf{Report evidence}\par
Running the \texttt{semgrep\_scan\_local} tool fails with:
\begin{tcolorbox}[colback=benchmarkexamplecode,colframe=benchmarkexampleframe,boxrule=0.3pt,arc=1pt,
  left=3pt,right=3pt,top=3pt,bottom=3pt,before skip=3pt,after skip=0pt]
\ttfamily\footnotesize Error executing tool semgrep\_scan\_local:\par
Error running semgrep scan: 'NoneType' object has no attribute 'decode'
\end{tcolorbox}
\end{tcolorbox}
\begin{tcolorbox}[enhanced,colback=white,colframe=benchmarkexampleframe,colbacktitle=benchmarkconfig,
  coltitle=black,fonttitle=\bfseries,title={(b) Case 0067: configuration artifacts},
  boxrule=0.5pt,arc=2pt,left=5pt,right=5pt,top=4pt,bottom=4pt,before upper=\raggedright]
\small
\textbf{Issue:} \href{https://github.com/rohitg00/kubectl-mcp-server/issues/61}{rohitg00/kubectl-mcp-server\#61}

\medskip
\textbf{Title}\par
\texttt{HTTPConnectionPool(host='localhost', port=80): Max retries exceeded}

\medskip
\medskip
\medskip
\textbf{Report evidence}\par
The reporter deployed the MCP server through a Kubernetes manifest and encountered the error when using it from an n8n MCP client. A transcription of the attached diagnostic image records a failed \texttt{get\_nodes} call over SSE and a refused connection to \texttt{/api/v1/nodes} on \texttt{localhost:80}.
\end{tcolorbox}
\end{tcbraster}
\caption{Examples of issue-derived task descriptions supplied to the repair agent. Case~0213 retains the reported tool failure and error message after correcting a typographical error in the tool name. Case~0067 combines the report text with a transcription of its diagnostic image. Neither description includes the developer repair or evaluator requirements.}
\Description{Two equal-height panels present examples of agent-visible task descriptions. The first reports a Semgrep tool failure caused by a missing standard-error value. The second reports a refused localhost connection after Kubernetes deployment and summarizes the diagnostic attachment.}
\label{fig:benchmark_task_examples}
\end{figure}

\subsubsection{Candidate case construction}

After Stage~A, we identify the buggy and developer-fixed repository revisions by their Git commit hashes. For each candidate case, we then define a \emph{behavioral contract} that specifies the externally observable requirements that a successful repair must satisfy. We derive these requirements from the issue report and eligible pre-fix discussion, reproduction steps, logs or attachments, relevant repository documentation and existing tests, the developer fix, and the behavior observed on the buggy and developer-fixed revisions. The developer fix provides evidence about the intended behavior, but it is not treated as the only acceptable implementation. When a repair may affect related behavior that already works, the contract also specifies what must be preserved.

Depending on the issue, the contract may concern capability discovery, accepted arguments, response structure, lifecycle or transport behavior, interactions with external services, or configuration and packaging. It translates the available issue and repair evidence into evaluation requirements without prescribing how the repository must be changed. Repairs using different SDK APIs, files, or repository structures can therefore satisfy the same contract through different edits.

We use these requirements to construct the case-specific evaluator and determine whether a candidate repair is successful. We also construct an offline execution environment for the case.
Existing benchmarks of real bugs similarly retain paired buggy and developer-fixed revisions and distinguish their behavior through executable tests or reproduction procedures~\cite{defects4j2014,manybugsIntroClass2015,bears2019,bugsinpy2020}. Each evaluator checks the behavior reported in the issue. When a repair may affect relevant existing behavior, the evaluator also includes \emph{preservation checks}. These checks can ensure, for example, that a repair does not remove a feature, hard-code one input, or fix one interface while breaking another. They reduce the established risk that incomplete tests accept plausible but behaviorally incorrect patches~\cite{qi2015,swebenchCorrectness2026}. A candidate repair can therefore pass without changing the same files or reproducing the developer's diff.

We use two evaluator forms. A \emph{runtime-behavior evaluator} executes the affected project code with controlled inputs. If the code normally communicates with a changing or credential-dependent service, such as a hosted API, the evaluator replaces only that service with a fixed local substitute while continuing to execute the project logic related to the issue. For example, the evaluator for benchmark case \texttt{case\_0212}, based on issue~\#25 in the \textit{Meta Ads MCP} repository, executes the project's creative-creation function while replacing the Meta Ads API with a local fake.\footnote{\url{https://github.com/pipeboard-co/meta-ads-mcp/issues/25}} The fake rejects the buggy request format and returns a fixed response for valid requests. The evaluator also checks that the existing dynamic-creative behavior remains supported~\cite{mcpRepairBenchArtifact}.

A \emph{configuration/package evaluator} checks the content and consistency of public artifacts such as configuration files, dependency manifests, Dockerfiles, and documentation when those artifacts determine the reported behavior. Prior repair research similarly treats configuration and other structured artifacts as valid repair targets~\cite{liu_llm_compdroid_2026,alloyRepair2025,terraRepair2026}.

Figure~\ref{fig:evaluator_examples} contrasts the two evaluator forms. Case~0213 executes the affected Python function with controlled subprocess outcomes, whereas case~0067 inspects the deployment and provider artifacts that determine the reported in-cluster configuration behavior.

\begin{figure}[htbp]
\centering
\begin{tcbraster}[raster columns=2,raster equal height=rows,raster column skip=6pt]
\begin{tcolorbox}[enhanced,colback=white,colframe=benchmarkexampleframe,colbacktitle=benchmarkruntime,
  coltitle=black,fonttitle=\bfseries,title={(a) Case 0213: runtime behavior},
  boxrule=0.5pt,arc=2pt,left=5pt,right=5pt,top=4pt,bottom=4pt,before upper=\raggedright]
\small
\textbf{Required behavior}\par
A failed scan with no \texttt{stderr} raises a stable \texttt{McpError}, rather than exposing an internal \texttt{NoneType.decode} exception.

\medskip
\medskip
\medskip
\textbf{Evaluation method}\par
Execute the affected scan function with controlled subprocess results for a missing diagnostic, a successful scan, and a failure with genuine \texttt{stderr}.

\medskip
\textbf{Preservation checks}\par
Successful scan output remains unchanged; genuine \texttt{stderr} diagnostics remain available.

\medskip
\textbf{Reference observations}\par
Buggy revision:
\textcolor{red!70!black}{\textbf{missing \texttt{stderr} causes a~\texttt{NoneType.decode} error}}\par
Fixed revision:
\textcolor{green!45!black}{\textbf{stable \texttt{McpError};
successful output and diagnostics preserved}}
\end{tcolorbox}
\begin{tcolorbox}[enhanced,colback=white,colframe=benchmarkexampleframe,colbacktitle=benchmarkconfig,
  coltitle=black,fonttitle=\bfseries,title={(b) Case 0067: configuration artifacts},
  boxrule=0.5pt,arc=2pt,left=5pt,right=5pt,top=4pt,bottom=4pt,before upper=\raggedright]
\small
\textbf{Required behavior}\par
The deployment selects the in-cluster provider and service port; configuration errors explain how to select that provider; and the provider can fall back to in-cluster configuration.

\medskip
\textbf{Evaluation method}\par
Inspect the deployment manifest and two provider/configuration modules for four required conditions.
\medskip

\medskip
\medskip
\textbf{Preservation checks}\par
No separate preservation scenario.

\medskip
\medskip
\medskip
\medskip
\textbf{Reference observations}\par
Buggy revision: \textcolor{red!70!black}{\textbf{required conditions absent}}\par
Fixed revision: \textcolor{green!45!black}{\textbf{all four conditions satisfied}}
\end{tcolorbox}
\end{tcbraster}
\caption{Examples of MCP-RepairBench evaluators. Case~0213 executes the affected function under one reported-failure and two preservation scenarios. Case~0067 checks four required conditions across deployment and provider artifacts without a separately evaluated preservation scenario. Both evaluators fail on the buggy revision and pass on the developer-fixed revision while permitting repairs that differ from the developer patch.}
\Description{Two equal-height panels summarize a runtime-behavior evaluator and a configuration/package evaluator using the same four-part structure: required behavior, evaluation method, preservation checks, and reference observations. The runtime case changes from an internal decoding error to a stable MCP error while preserving successful output and genuine diagnostics. The configuration case has no separately evaluated preservation scenario and changes from absent required conditions to all four conditions being satisfied.}
\label{fig:evaluator_examples}
\end{figure}

We record two versions of the developer repair for each task. The \emph{complete developer-fix scope} contains every file changed by the linked repair, whereas the \emph{issue-specific scope} contains only the implementation, test, configuration, or documentation changes associated with the focal issue. In nine cases, the complete repair includes unrelated changes that are excluded from the issue-specific scope (see Appendix~\ref{subsec:fix_test_scope} for an example). Neither scope is shown to the repair agent or used as an answer key. The evaluator determines whether a generated repair is successful.

\subsubsection{Stage B: case acceptance review}
The completed candidate case then undergoes Stage~B using the full construction record, including evidence not supplied to the repair agent. The reviewer records five judgments concerning issue and fix validity, contract fairness, evaluator validity, applicable preservation requirements, and license adequacy. These judgments assess the selected revisions, patch scope, behavior observed on the buggy and fixed revisions, incomplete or shared repairs, and the execution environment. This process follows established benchmark-verification practices for assessing issue clarity, evaluator appropriateness, and the validity of issue-resolving cases~\cite{swebenchverified2024,multiswebench2025}.

Stage~B excludes two candidates. In one case, executing the completed evaluator shows that the selected fixed revision does not establish the required behavior. The other requires the same underlying repair as a retained candidate and therefore does not form an independent task. Their replacements repeat the process shown in Figure~\ref{fig:benchmark_construction_pipeline}. Appendix~\ref{appendix} reports the complete review questions, evidence rules, decision process, and stage-specific replacement reasons.

\subsubsection{Final release verification}
After Stage~B acceptance, we build each task's offline environment. To test repeatability, we execute its evaluator five times on both the buggy and developer-fixed states with networking disabled; RQ3 reports the resulting 600 verdicts. Appendix~\ref{appendix} records the release-verification evidence. The released benchmark contains 60 tasks, including 44 from the initial selection and 16 replacements. Of these tasks, 48 use runtime-behavior evaluators and 12 use configuration/package evaluators.

\subsubsection{Final task structure and scoring}
\begin{figure}[htbp]
\centering
\includegraphics[width=0.8\textwidth]{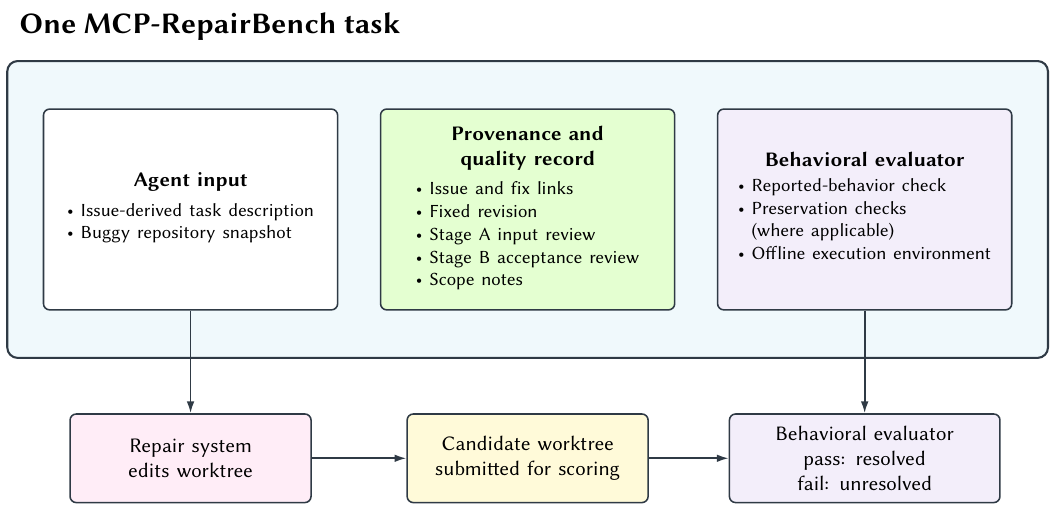}
\caption{Structure and evaluation of an MCP-RepairBench task. Each task combines the input supplied to a repair system with provenance and review records and a behavioral evaluator that are withheld from it. The repair system modifies the buggy worktree, and the resulting candidate is evaluated against the reported behavior and, where applicable, behavior designated for preservation.}
\Description{A container groups three parts of a benchmark task, including repair-system input, provenance and quality records, and a behavioral evaluator. The provenance records and evaluator remain hidden from the repair system. Below the container, a straight evaluation path runs from the repair system through a candidate worktree to the evaluator, whose checks cover the reported behavior and any applicable preservation requirements. Its pass or fail verdict determines whether the task is resolved.}
\label{fig:benchmark_task_schema}
\end{figure}

The finalized MCP-RepairBench follows established software-repair benchmark practices by linking real bug reports to reproducible buggy and developer-fixed revisions and evaluators~\cite{defects4j2014,bugsinpy2020,swebench2024}. Each task combines its reviewed task description, buggy and developer-fixed repository revisions, behavioral evaluator, offline execution environment, and provenance and review records (Figure~\ref{fig:benchmark_task_schema}).

During evaluation, a repair system receives the task description and a writable worktree, which is an isolated, modifiable copy of the repository at the buggy revision. The evaluator and reference materials remain inaccessible to the system. The repair system may modify any file in the worktree. A task is \emph{resolved} only when its evaluator confirms the reported behavior and any behavior designated for preservation. Compilation, startup, timeout, and evaluator failures are counted as unresolved.

\subsection{MCP-aware repair agent and evaluation}\label{subsec:agent_evaluation}

RQ4 evaluates the MCP-aware repair agent on the finalized MCP-RepairBench constructed in RQ3. We first present the agent's structure and execution flow, followed by the baseline design, experimental conditions and evaluation protocol, and outcome measures and statistical analysis.

\begin{figure}[htbp]
\centering
\includegraphics[width=\textwidth]{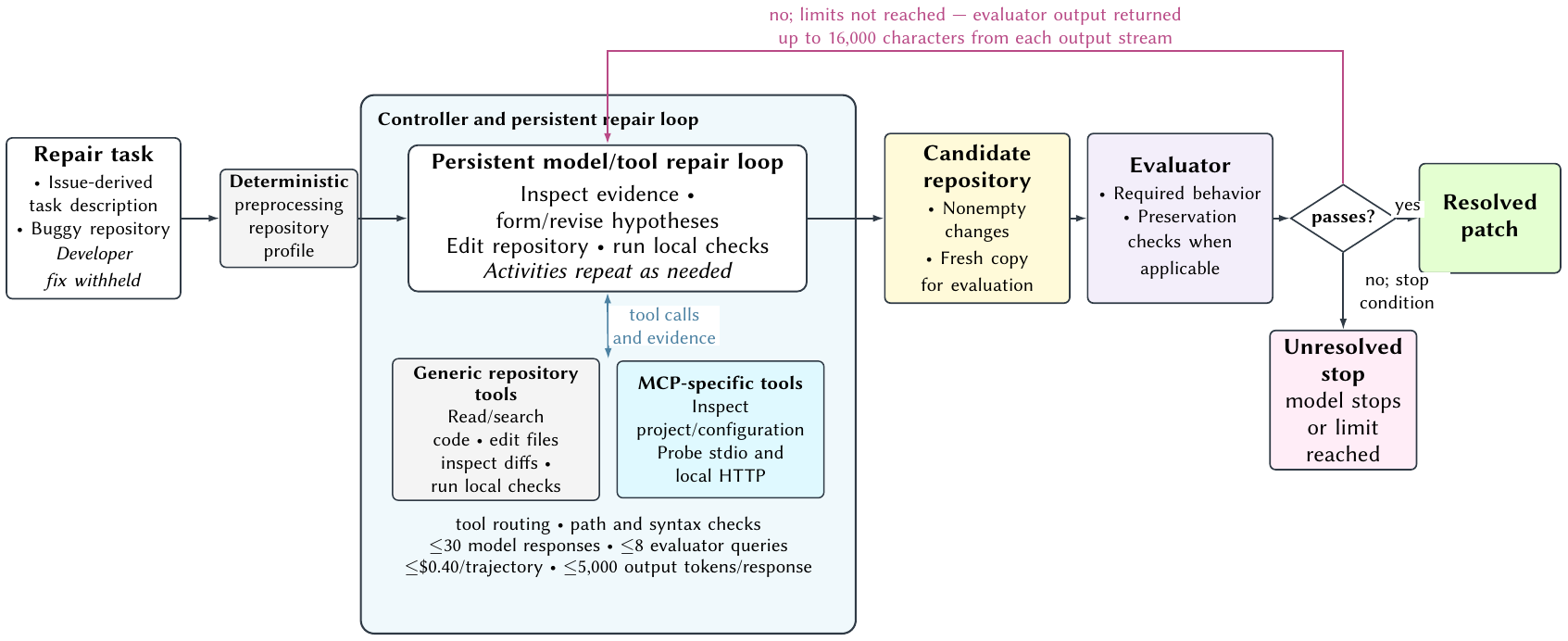}
\caption{Evaluated MCP-aware repair agent. The controller first constructs the deterministic repository profile, shown separately as preprocessing before the persistent model/tool loop. The loop flexibly interleaves inspection, hypothesis formation, editing, and local checks. Blue denotes interaction with the generic and MCP-specific tool palette. Pink denotes the recurrent evaluator-feedback path, which may be traversed for up to eight changed candidates per trajectory. The agent configuration, tools, limits, and stopping rules were fixed before evaluation.}
\Description{Left-to-right repair-agent architecture and execution process. An issue-derived task description and buggy repository enter deterministic preprocessing, through which the controller constructs a repository profile before model interaction. The profile is supplied to a controller containing one persistent model/tool loop. Inspection, hypothesis formation, editing, and local checks can recur in any order, supported by generic repository and MCP-specific tools. A repository containing nonempty changes is copied to a fresh evaluation directory and checked by an evaluator. A passing candidate yields a resolved patch. After a failure, evaluator output returns to the repair loop if no stopping limit has been reached; otherwise, the trajectory ends unresolved because the model stops or a limit is reached.}
\label{fig:agent_workflow}
\end{figure}

\paragraph{Agent structure and execution flow.}
The MCP-aware repair agent represents an initial step toward automated repair of MCP bugs. We use \emph{agent} to denote the complete model--harness configuration, which includes the model, controller, prompts, repository and MCP-specific tools, context-management policy, evaluator-feedback interface, budgets, and stopping rules. Figure~\ref{fig:agent_workflow} depicts the agent within the evaluated workflow. The agent combines deterministic repository profiling, a controller that maintains one persistent model--tool loop, and generic and MCP-specific tools exposed through the harness. Its feedback interface connects the controller to a behavioral evaluator that executes separately from the agent.

A \textit{trajectory} comprises all model interactions, tool actions, and environment observations produced during one independently initialized repair attempt~\cite{sweagent2024}. Each trajectory begins with the issue-derived task description and the corresponding buggy version of the repository. A deterministic profiler identifies programming languages from file extensions, records common project and build files, finds files containing explicit MCP SDK markers, proposes generic local checks based on the detected project type, and produces a compact repository file listing.
The controller supplies this profile to the LLM and maintains one continuous conversation throughout the trajectory. Across successive responses, the model can inspect evidence, formulate and revise hypotheses, edit the repository, and perform local checks in any order; these activities are not enforced as separate workflow stages. This flexible, tool-mediated interaction follows established repository-level repair-agent designs~\cite{repairagent2024,sweagent2024}. When a nonempty changed candidate is submitted, the controller sends a fresh copy of the candidate repository to the isolated evaluator. If evaluation fails and the trajectory has not reached a stopping limit, the controller returns the exit code and up to the final 16,000 characters from each of the evaluator's standard-output and standard-error streams. If either stream is longer, its earlier text is omitted. The controller replaces the candidate-workspace path and absolute paths explicitly included in the evaluator command with placeholders. Other content, including assertion messages, observed values, and traceback frames, remains available to the model. Appendix~\ref{subsec:evaluator_feedback_example} provides an example. This feedback loop follows prior conversational and agentic repair systems in using validation results to refine successive candidates~\cite{chatrepair2024,engineeringAgent2026}.

Table~\ref{tab:rq4_tools} details the tool interface. Generic tools support repository inspection, local execution, and editing, while MCP-specific tools provide focused evidence about server construction, schemas, discovery, transports, protocol behavior, and configuration. We derived these capabilities from recurring repair needs observed among the 197 coded repairs in Phase~I. Schema and registration repairs motivated inspection of MCP server construction and registered tools. Startup and configuration repairs motivated tracing of environment variables, defaults, and launch settings. Transport, response-handling, and protocol-output repairs motivated local \texttt{stdio} and HTTP probes.
The model receives the available tools together with descriptions of their functions and intended uses. Throughout the trajectory, it decides which tool, if any, to invoke. The controller does not select tools on the model's behalf. No Phase~I issue, label, or developer patch is supplied to the agent during evaluation. Similar use of domain-specific evidence tools appears in prior repair-agent research~\cite{gradleFixer2026,vulDebugger2025,terraRepair2026}. 
Before benchmark evaluation, we finalized the tool interface and experimental settings using an eight-case \emph{development suite} derived from Phase~I. None of these cases appears in MCP-RepairBench. Details of the development cases and their experimental records are available in the repair-agent replication package~\cite{mcpRepairAgentArtifact}.

\begin{table}[htbp]
\scriptsize
\caption{Tools available to the MCP-aware repair agent. Generic tools support repository inspection, execution, and editing. MCP-specific tools inspect and exercise MCP-related project behavior.}
\label{tab:rq4_tools}
\centering
\setlength{\tabcolsep}{4pt}
{%
\renewcommand{\tabularxcolumn}[1]{m{#1}}
\renewcommand{\arraystretch}{1.08}
\begin{tabularx}{\textwidth}{>{\raggedright\arraybackslash}m{0.105\textwidth}>{\raggedright\arraybackslash}m{0.222\textwidth}>{\raggedright\arraybackslash}m{0.326\textwidth}>{\raggedright\arraybackslash}X}
\toprule
\textbf{Type} & \textbf{Tool} & \textbf{What it does} & \textbf{What the model receives} \\ \midrule
\rowcolor{black!4}
Generic & \texttt{read\_file} & Reads a requested range of numbered lines from a specified repository file & The requested lines and their line numbers \\
Generic & \texttt{search\_code} & Searches repository text files for a literal string or regular expression, optionally within matching file paths & Matching passages with file paths, line numbers, and nearby lines \\
\rowcolor{black!4}
Generic & \texttt{find\_symbol} & Searches for likely definitions and references of a named class, function, method, or identifier & Matching definition and reference passages \\
Generic & \texttt{run\_command} & Runs one approved compile, lint, test, build, or diagnostic program from the repository directory & Exit code, standard output, and standard error \\
\rowcolor{black!4}
Generic & \texttt{replace\_text} & Replaces one exact, uniquely matching text block in an existing repository file & Edit confirmation and the automatic Python syntax-check result \\
Generic & \texttt{create\_file} & Creates a new repository file with the supplied contents & Creation confirmation and the automatic Python syntax-check result \\
\rowcolor{black!4}
Generic & \texttt{apply\_patch} & Applies a supplied patch to one or more repository files & Patch result and the automatic Python syntax-check result \\
Generic & \texttt{git\_diff} & Reads all current tracked and newly created changes in the repository & The current candidate patch \\ \midrule
\rowcolor{black!4}
MCP-specific & \texttt{inspect\_mcp\_project} & Examines the repository for MCP server construction, tool registration, transports, configuration access, and protocol-output code & Ranked MCP-related files and matching evidence for each inspected area \\
MCP-specific & \texttt{inspect\_mcp\_configuration} & Traces a supplied configuration term and common environment-variable and command-line configuration APIs across the repository & Matching definitions and uses grouped by search term \\
\rowcolor{black!4}
MCP-specific & \texttt{probe\_mcp\_stdio} & Starts a local MCP server over stdio, initializes it, and optionally lists or invokes a tool & Parsed protocol messages, non-protocol output, errors, exit status, and timeout status \\
MCP-specific & \texttt{probe\_mcp\_http} & Starts a local MCP server and sends one supplied GET, POST, or OPTIONS request to a localhost endpoint & HTTP status, response headers, and response content \\ \midrule
\rowcolor{black!4}
Evaluation & \texttt{evaluate\_candidate} & Executes the current changed repository with the benchmark evaluator & Pass or fail, exit code, evaluator output, and whether the same failure occurred previously \\ \bottomrule
\end{tabularx}
}%
\end{table}

The harness is the execution layer that carries out the model's tool requests within the task repository and returns their results. It resolves file paths against the repository root and rejects absolute paths or attempts to access a parent directory. For local checks, the model supplies a program and its arguments as a structured list. The harness accepts only a fixed set of common test and build programs, such as \texttt{python}, \texttt{pytest}, \texttt{uv}, \texttt{npm}, and \texttt{make}, and launches the selected program directly from the repository directory without exposing a general-purpose shell. Shell syntax for chaining, piping, or redirecting commands is therefore not interpreted. After each edit, the harness uses the Python compiler to check changed \texttt{.py} files for syntax and encoding errors without executing them. It immediately returns any errors to the model so that malformed edits can be corrected before candidate evaluation. The complete list of allowed programs and harness configuration are recorded in the replication package~\cite{mcpRepairAgentArtifact}.

The controller rejects an evaluation request when the repository contains no changes or when the current diff has already been evaluated. The model may explicitly request evaluation of its current patch. If an edited patch remains unevaluated for four model responses, the controller evaluates it automatically. It also evaluates any outstanding patch before the trajectory ends. Each candidate is evaluated in a disposable container with network access disabled. Before repository tools run, the LLM-provider client is initialized and its API key is removed from the process environment; no external-service credentials are supplied to the repository or evaluator.

\paragraph{Continuous-loop baseline.}
We use a generic repair baseline because repository-level repair agents commonly alternate between inspecting the repository, editing files, and running available local tests or diagnostic commands~\cite{sweagent2024,repairagent2024}. The baseline represents this common repair workflow rather than reproducing a specific published system. Prior work also shows that a relatively simple repair pipeline can perform competitively with more elaborate repair agents~\cite{agentless2024}. The baseline receives the same issue-derived task description and a worktree initialized from the same buggy repository revision as the MCP-aware agent, but retains a simpler, continuous inspection-and-editing loop. It exposes generic repository tools and withholds evaluator feedback until the final patch is scored. Compared with the MCP-aware agent, it therefore omits the MCP-specific evidence tools, controller-managed candidate evaluations, and recurrent evaluator-feedback path. The MCP-aware agent and the baseline use the same per-trajectory resource ceilings, reported below.

\paragraph{Experimental conditions.}
We evaluate the MCP-aware agent and continuous-loop baseline on all 60 MCP-RepairBench tasks using the pinned \texttt{gpt-5.4-mini-2026-03-17} snapshot, a cost-efficient model intended for coding and tool use~\cite{openai_gpt54mini_model}. Using this development suite, we selected the model provider’s \texttt{``low''} reasoning-effort setting as a balance between repair performance and API cost. We fixed this setting before benchmark evaluation and used it for every GPT-5.4-mini trajectory.
To assess sensitivity to model choice, we also evaluate the MCP-aware agent with the pinned \texttt{gpt-4.1-mini-2025-04-14} snapshot, an earlier-generation model with documented instruction-following and tool-calling capabilities~\cite{openai_gpt41mini_model}. The two models provide a deliberately contrasting condition. GPT-5.4-mini supports configurable reasoning and has an August~2025 knowledge cutoff, whereas GPT-4.1-mini is a non-reasoning model whose June~2024 cutoff precedes Anthropic's public introduction of MCP on November~25, 2024~\cite{openai_gpt54mini_model,openai_gpt41mini_model,anthropicMCPLaunch2024}. The reasoning-effort field is therefore omitted for GPT-4.1-mini. This contrast tests whether the MCP-aware agent remains useful with an older model and knowledge horizon. Sampling parameters remain at their API defaults. We do not specify temperature, top-\(p\), or seed in any condition. It should be taken into account that GPT-5.4-mini does not support temperature or top-\(p\) at the selected reasoning setting~\cite{openai_gpt54mini_model}; for GPT-4.1-mini, we leave these parameters unspecified to maintain the same default-sampling policy.

The MCP-aware agent and continuous-loop baseline use predeclared limits to keep their trajectories finite and their resource use auditable, as is common in repository-level repair-agent evaluations~\cite{repairagent2024,sweagent2024}. Each trajectory permits at most 30 model responses, 30 serial model-invoked tool calls, 5,000 output tokens per response, an API-cost ceiling of \$0.40, and 120 seconds for each local command. The MCP-aware configuration permits at most eight evaluator queries per trajectory; the baseline receives no evaluator feedback during repair. Earlier development runs allowed 18 responses and three evaluator queries. We increased these limits to 30 responses and eight queries because some development trajectories reached the smaller limits while their candidate repairs were still improving. The 5,000-token, cost, and command-time ceilings were set above the use observed during development to prevent unusually long responses, excessive cost, and stalled commands. We did not optimize these three limits separately. The 30-response limit is comparable to the 40-cycle default used by RepairAgent and the approximately 30--40 turns reported for SWE-agent~\cite{repairagent2024,sweagent2024}. These limits remain unchanged across fresh trajectories and, for the MCP-aware agent, across the two models. Because the agent and the baseline still differ in their prompts, controllers, tool interfaces, and feedback access, the matched-resource comparison concerns the complete repair systems rather than any single component~\cite{wohlin2024}. Logs record model inputs and outputs, tool events, patches, evaluator queries, token usage, API cost, elapsed time, and final status. Each configuration and its retry rules were fixed before evaluation; the replication package records the complete configuration~\cite{mcpRepairAgentArtifact}.

\paragraph{Outcome measures and analysis.}
Within a trajectory, the MCP-aware agent may produce several changed candidates. Each execution of a candidate by the task evaluator is an \emph{evaluator query}. A failed query returns the evaluator output described above to the same trajectory, allowing the agent to revise its patch; a passing query resolves the task.

We execute one trajectory per task to obtain pass@1. Tasks that remain unresolved receive up to two additional independently initialized trajectories, and no further trajectory is executed after success. Under the standard pass@$k$ definition, a task passes if at least one of $k$ independent samples succeeds~\cite{chenCodeModels2021}; repair-agent evaluations apply the same definition to independent trajectories~\cite{sweagent2024,swebenchlive2025}. Accordingly, pass@3 is the fraction of tasks resolved in at least one of the three available trajectories.
Early stopping avoids unnecessary trajectories without changing this measure because a task already resolved remains a success.

For every condition, we report the number and proportion of the 60 benchmark tasks resolved. At pass@\(k\), the numerator is the number of tasks resolved in at least one of the available trajectories, and the denominator remains 60 because early stopping affects only the number of trajectories executed.

Because each pair of conditions is evaluated on the same 60 tasks, the results are paired by task. For each comparison, we report how many tasks are resolved by both conditions, only by the first condition, only by the second, or by neither. The exact McNemar test focuses on the tasks resolved by only one of the two conditions. It tests whether the imbalance between these two counts is consistent with equal resolution probabilities for the paired conditions~\cite{mcnemar1947}.

We also report results by runtime-behavior and configuration/package evaluator form and by benchmark coverage group, and examine execution traces and resource use. For the first-trajectory resource measures, we compare resolved and unresolved trajectories using two-sided Mann--Whitney U tests and apply Bonferroni correction across the four measures. The remaining subgroup and trace analyses are descriptive. API cost is calculated from the provider-reported counts of uncached input, cached input, and output tokens using the prices in effect during evaluation~\cite{openai_gpt54mini_model,openai_gpt41mini_model}. It excludes local execution, storage, and human effort.

\FloatBarrier

\section{Results}
\label{sec:result}
\raggedbottom
This section presents the findings of the study organized around the research questions (RQs).
RQ1 and RQ2 characterize the verified developer fixes from Phase~I, while RQ3 and RQ4 report the benchmark-construction and agent-evaluation evidence from Phase~II. Each RQ subsection concludes with a summary of its key findings.

\subsection{RQ1: Fix Patterns in MCP Bugs}\label{subsec:fix_patterns}
RQ1 reports the repair taxonomy derived from the 197 MCP-related issues 
retained through the procedure described in Section~\ref{subsec:semantic_repair_coding}. We first examine their distribution across repair groups and then the more specific action patterns within those groups. Because the taxonomy excludes verified fixes without a sufficiently localizable repair interpretation, its frequencies characterize the coded subset rather than all 221 fix-bearing issues.

The final taxonomy contains 10 repair groups and 60 concrete action patterns. The groups locate the affected repair boundary, whereas the actions describe the change developers made at that boundary. Table~\ref{tab:rq1_group_distribution} shows that \emph{Startup and Environment} is the largest group (22.3\%), followed by \emph{Tool Execution} (16.2\%), \emph{Tool Schema and Discovery} (14.7\%), and \emph{Documentation} (14.2\%). Together, these four groups account for 133 of 197 coded repairs (67.5\%). This concentration of repairs in the four largest groups is consistent with MCP's integration role. Repairs in these groups commonly make servers start in the intended environment, make tool calls execute correctly, align exposed schemas with host expectations, or correct setup guidance.

\begin{table}[h]
\small
\caption{Repair-group distribution in the 197 coded repairs. Percentages are rounded independently.}
\label{tab:rq1_group_distribution}
\centering
\begin{tabular}{lrr}
\toprule
\textbf{Repair group} & \textbf{Count} & \textbf{Percentage} \\ \midrule
Startup \& Environment & 44 & 22.3\% \\
Tool Execution & 32 & 16.2\% \\
Tool Schema \& Discovery & 29 & 14.7\% \\
Documentation & 28 & 14.2\% \\
External Service Integration & 19 & 9.6\% \\
Connection \& Transport & 13 & 6.6\% \\
Tool Response Handling & 13 & 6.6\% \\
Logging \& Protocol Hygiene & 8 & 4.1\% \\
Platform \& Build Compatibility & 6 & 3.0\% \\
Runtime State Management & 5 & 2.5\% \\ \midrule
Total & 197 & 100.0\% \\ \bottomrule
\end{tabular}
\end{table}

Table~\ref{tab:rq1_action_distribution} combines the most frequent actions with their repair groups. Tool-signature correction is the most common individual pattern, followed by installation-guidance clarification, dependency-version updates, and configuration-example corrections. The action distribution is nevertheless fragmented. Nineteen patterns occur once and 30 occur at most twice. We therefore use the better-populated groups for the statistical comparisons in RQ2 and retain all 60 actions as the fine-grained codebook reported in Appendix~\ref{subsec:rq1_codebook}.

\begin{table*}[h]
\small
\caption{Most frequent action patterns in the 197 coded repairs. The complete codebook appears in Appendix~\ref{subsec:rq1_codebook}.}
\label{tab:rq1_action_distribution}
\centering
\begin{tabular}{>{\raggedright\arraybackslash}p{0.38\textwidth}
    >{\raggedright\arraybackslash}p{0.27\textwidth}
    r}
\toprule
\textbf{Action pattern} & \textbf{Repair group} & \textbf{Count} \\ \midrule
Tool signature correction & Tool Schema \& Discovery & 15 \\
Installation guidance clarification & Documentation & 10 \\
Configuration example correction & Documentation & 9 \\
Dependency version update & Startup \& Environment & 8 \\
Startup configuration handling & Startup \& Environment & 8 \\
Startup path/resource setup & Startup \& Environment & 8 \\
Tool registration correction & Tool Schema \& Discovery & 7 \\
Provider configuration management & External Service Integration & 7 \\
Startup/runtime configuration correction & Startup \& Environment & 5 \\
Tool name correction & Tool Schema \& Discovery & 5 \\
Endpoint/path correction & Connection \& Transport & 5 \\
Request/endpoint correction & External Service Integration & 5 \\
Other 48 action patterns & All 10 groups & 105 \\ \midrule
\multicolumn{2}{l}{Total} & 197 \\ \bottomrule
\end{tabular}
\end{table*}

Three examples illustrate how the action level preserves mechanisms that group counts alone would obscure. First, schema incompatibilities in different implementations receive the same \emph{tool-signature correction} code when developers add missing array-item definitions or replace host-incompatible union schemas.\footnote{Examples: \url{https://github.com/meilisearch/meilisearch-mcp/issues/21} and \url{https://github.com/Rudra-ravi/wikipedia-mcp/issues/13}.} Second, moving a dashboard launch outside the stdio process is coded as \emph{protocol output isolation} because the repair prevents browser output from contaminating JSON messages.\footnote{Issue and fix: \url{https://github.com/oraios/serena/issues/257} and \url{https://github.com/oraios/serena/commit/80a4caf36a1692c71888bb92ab0f03ebe7637c73}.} Third, correcting malformed JSON in a documented host configuration is a \emph{configuration example correction}. Although it changes documentation rather than source code, the published artifact is the setup interface that users execute.\footnote{Issue and fix: \url{https://github.com/awslabs/mcp/issues/526} and \url{https://github.com/awslabs/mcp/pull/527}.}

\begin{keyfindings}{RQ1}
\begin{keyfindingslist}
    \item The 197 localizable repairs form 10 repair groups and 60 action patterns.
    \item \emph{Startup and Environment}, \emph{Tool Execution}, \emph{Tool Schema and Discovery}, and \emph{Documentation} account for 67.5\% of the coded repairs, concentrating repair work at a small number of integration boundaries.
    \item No individual action dominates. The most frequent occurs in 15 repairs, and 30 of the 60 action patterns occur at most twice.
\end{keyfindingslist}
\end{keyfindings}

\subsection{RQ2: Characteristics of MCP Bug Fixes}\label{subsec:fix_characteristics}
RQ2 characterizes the complete set of 221 fix-bearing issues and then examines how those measurements vary across the repair groups established in RQ1. At the issue level, 148 fix-bearing issues are represented by PRs and 73 by direct commits. Five issues contain two genuine fix artifacts, yielding 226 artifacts in total, including 152 pull requests and 74 commits. Table~\ref{tab:rq2_summary} summarizes the repair metrics for the 221 fix-bearing issues. Added and deleted LOC are included only to show the composition of changed LOC; they are not treated as separate metrics in the repair-group comparisons. The fixes change a median of two files and 59 LOC. Both distributions contain a small number of much larger fixes. One fix changes 121 files, and nine fixes change more than 3,000 LOC.

\begin{table}[h]
\small
\caption{Issue-level summary of fix characteristics ($N=221$ except where shown). Completion latency excludes the 11 issues whose fix completion predates the issue report ($N=210$); PR duration is available for the 148 issues represented by a merged PR.}
\label{tab:rq2_summary}
\centering
\begin{tabular}{lrrr}
\toprule
\textbf{Metric} & \textbf{Median} & \textbf{Mean} & \textbf{Max} \\ \midrule
Changed files & 2.0 & 6.41 & 121 \\
Changed LOC & 59.0 & 611.36 & 13,134 \\
Added LOC & 38.0 & 436.79 & 11,582 \\
Deleted LOC & 7.0 & 174.57 & 6,839 \\
Commits in fix & 1.0 & 3.31 & 32 \\
Fix completion latency (hours; $N=210$) & 49.94 & 194.66 & 2,342.06 \\
Merged-PR duration (hours; $N=148$) & 3.61 & 32.88 & 610.55 \\
Change entropy & 0.82 & 1.07 & 5.43 \\ \bottomrule
\end{tabular}
\end{table}

Seventy-nine fixes change one file, and 117 change no more than two files. In addition, 102 fixes change no more than 50 LOC, and 65 are both single-file and no larger than 50 LOC. These counts show that many repairs are localized. A smaller number of much larger fixes, including dependency migrations, generated-code updates, and changes spanning several parts of a project, raise the means well above the medians. The medians therefore better describe the size of the more common repairs, while the larger cases show the range that repair tools and benchmarks must also support.

We examine all 20 correlations between metrics from different families. Table~\ref{tab:rq2_cross_family_correlations} reports the results. 
Three remain significant after Bonferroni correction. Changed LOC, file count, and entropy are positively correlated with completion latency ($\rho=0.22$--$0.26$), whereas commit count is not. None of the structural metrics is significantly correlated with merged-PR duration. Neither repository nor global fixer experience is significantly correlated with the structural or process metrics. These results suggest that larger or more dispersed fixes tend to have somewhat longer intervals between issue creation and fix completion. They do not show a corresponding increase in PR duration or a consistent relationship with fixer experience.

\begin{table}[t]
\footnotesize
\caption{Spearman correlations between metrics from different families. Comparisons of structural metrics with completion latency, merged-PR duration, and fixer experience use 210, 148, and 196 fixes, respectively. Process--experience comparisons use 186 fixes for completion latency and 131 for merged-PR duration. Bold values remain significant at $p<0.05$ after Bonferroni correction across all 20 tests.}
\label{tab:rq2_cross_family_correlations}
\centering
\setlength{\tabcolsep}{5pt}
\renewcommand{\arraystretch}{1.3}
\begin{tabular}{clrrrr}
\toprule
\multirow{2}{*}{\textbf{Family}} & \multirow{2}{*}{\textbf{Metric}} & \multicolumn{2}{c}{\textbf{Process metrics}} & \multicolumn{2}{c}{\textbf{Fixer experience}} \\
\cmidrule(lr){3-4}\cmidrule(lr){5-6}
& & {Completion} & {PR duration} & {Repository} & {Global} \\ \midrule
\multirow{4}{*}{\textit{Structural metrics}}
& Changed LOC & \textbf{0.259} & 0.015 & 0.177 & $-0.123$ \\
& Changed files & \textbf{0.241} & 0.050 & 0.209 & $-0.029$ \\
& Change entropy & \textbf{0.217} & 0.025 & 0.191 & $-0.037$ \\
& Commits in fix & 0.131 & 0.176 & 0.026 & $-0.107$ \\ \midrule
\multirow{2}{*}{\textit{Process metrics}}
& Completion latency & -- & -- & $-0.040$ & $-0.173$ \\
& Merged-PR duration & -- & -- & $-0.224$ & $-0.096$ \\ \bottomrule
\end{tabular}
\end{table}
\FloatBarrier

\subsubsection{Differences across repair groups}
\begin{figure}[!tb]
\centering
\subfloat[]{\includegraphics[width=0.48\textwidth]{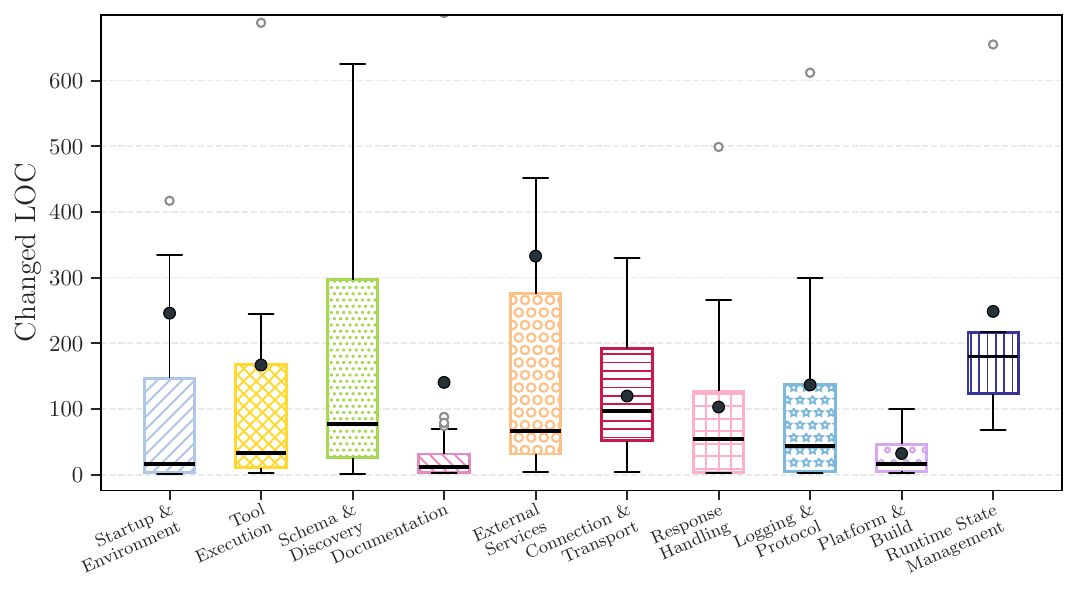}}
\subfloat[]{\includegraphics[width=0.48\textwidth]{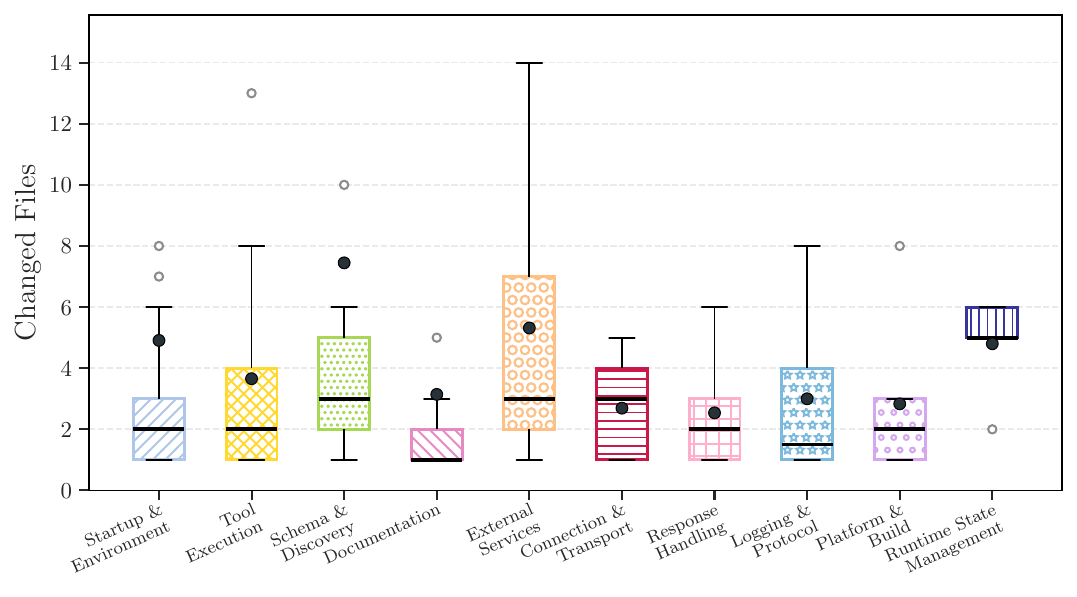}}\\
\subfloat[]{\includegraphics[width=0.48\textwidth]{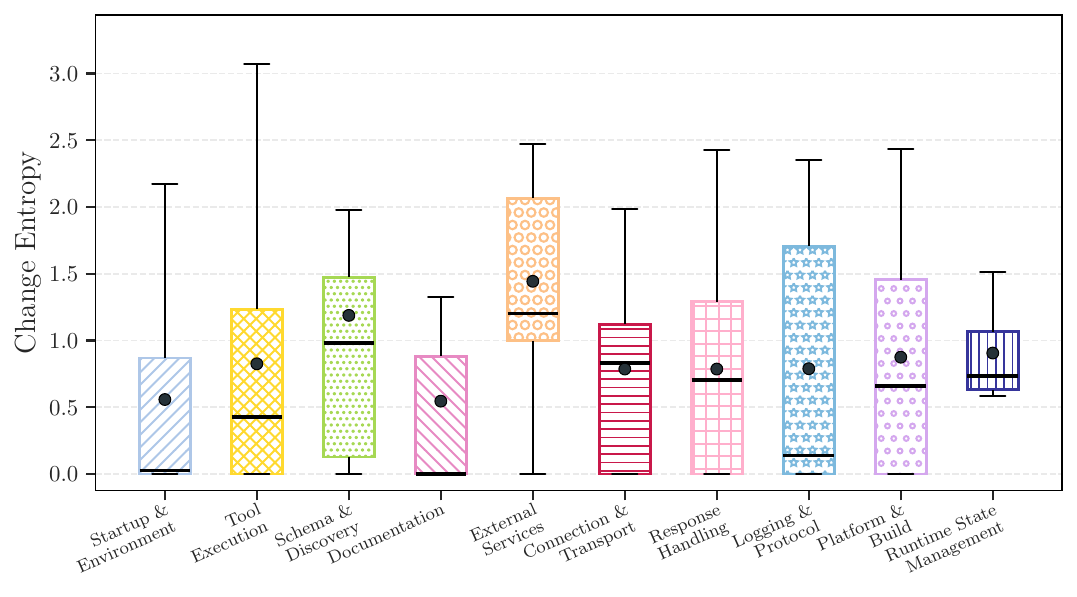}}
\subfloat[]{\includegraphics[width=0.48\textwidth]{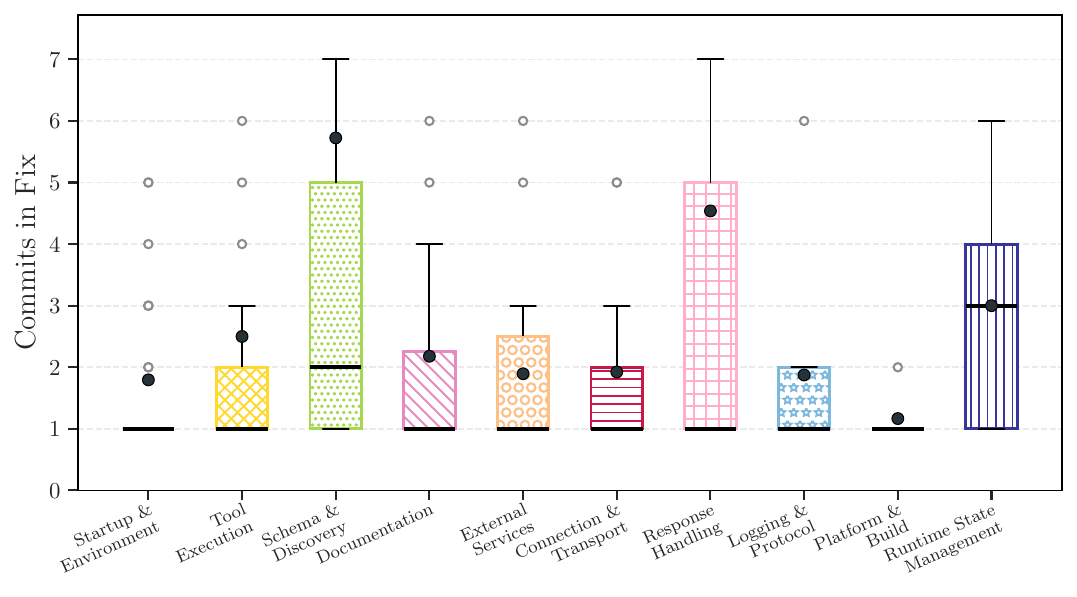}}
\caption{Structural fix metrics across the 10 repair groups. The panels show (a) changed LOC, (b) changed files, (c) change entropy, and (d) commits in the fix. Dark filled circles indicate means and open grey circles indicate outliers. Extreme outliers outside the displayed ranges are clipped.}
\Description{Four boxplots compare changed lines, changed files, entropy, and commit count across ten MCP repair groups. Documentation tends to have smaller, more localized fixes, whereas repairs in Tool Schema and Discovery, External Service Integration, and Runtime State Management tend to be broader.}
\label{fig:repair_group_metrics}
\end{figure}

Table~\ref{tab:rq2_group_medians} and Figure~\ref{fig:repair_group_metrics} summarize the four structural metrics. Table~\ref{tab:rq2_group_kruskal} reports Kruskal--Wallis results for all eight metrics, and Table~\ref{tab:rq2_group_dunn} reports the significant Bonferroni-adjusted comparisons. We interpret the descriptive and statistical results together; a high or low median alone is not treated as evidence of a reliable group difference.

\begin{table}[!tbp]
\small
\caption{Median structural fix characteristics across the repair groups. Bold marks the maximum and unique minimum in each metric column; the minimum commit median (1.0) is shared by eight groups and is not emphasized.}
\label{tab:rq2_group_medians}
\centering
\begin{tabular}{lrrrrr}
\toprule
\textbf{Repair group} & \textbf{$N$} & \textbf{LOC} & \textbf{Files} & \textbf{Entropy} & \textbf{Commits} \\ \midrule
Startup \& Environment & 44 & 16.5 & 2.0 & 0.03 & 1.0 \\
Tool Execution & 32 & 32.5 & 2.0 & 0.43 & 1.0 \\
Tool Schema \& Discovery & 29 & 77.0 & 3.0 & 0.98 & 2.0 \\
Documentation & 28 & \textbf{12.0} & \textbf{1.0} & \textbf{0.00} & 1.0 \\
External Service Integration & 19 & 67.0 & 3.0 & \textbf{1.20} & 1.0 \\
Connection \& Transport & 13 & 97.0 & 3.0 & 0.83 & 1.0 \\
Tool Response Handling & 13 & 54.0 & 2.0 & 0.71 & 1.0 \\
Logging \& Protocol Hygiene & 8 & 44.0 & 1.5 & 0.14 & 1.0 \\
Platform \& Build Compatibility & 6 & 16.0 & 2.0 & 0.66 & 1.0 \\
Runtime State Management & 5 & \textbf{180.0} & \textbf{5.0} & 0.73 & \textbf{3.0} \\ \bottomrule
\end{tabular}
\end{table}

\paragraph{Structural differences.}
Figure~\ref{fig:repair_group_metrics} shows considerable variation within each repair group, including a small number of comparatively large fixes. The Kruskal--Wallis tests show statistically significant differences of medium magnitude across groups for changed LOC, files, entropy, and commits (Table~\ref{tab:rq2_group_kruskal}).

The pairwise comparisons identify the specific repair-group differences. \emph{Tool Schema and Discovery} fixes change more LOC and files than \emph{Documentation} fixes and require more commits than \emph{Startup and Environment} fixes (Table~\ref{tab:rq2_group_dunn}). A tool-schema or registration repair may require coordinated changes to the tool schema presented to the MCP client, the server code that registers the tool, and the function that executes tool requests. By contrast, a documentation repair may require changing only a configuration example or setup instruction in the repository documentation. RQ1 likewise identifies tool-signature and tool-registration corrections as recurring action patterns within \emph{Tool Schema and Discovery}. The more localized documentation repairs also align with Taraghi et al.'s observation that documentation issues generally require less diagnostic interaction than server--host configuration faults~\cite{ourPaper}.

\emph{External Service Integration} has the highest median change entropy (1.20). The post-hoc comparisons show that its fixes have higher change entropy than \emph{Documentation} and \emph{Startup and Environment} fixes. They also modify more files than \emph{Documentation} fixes (Table~\ref{tab:rq2_group_dunn}). RQ1 provides a possible explanation for this result. Twelve of the 19 \emph{External Service Integration} fixes involve provider-configuration management or request/endpoint correction (Table~\ref{tab:rq1_action_distribution}). These fixes can change the code that configures or calls an external provider together with related configuration and tests, which can increase both the number of changed files and change entropy.

\begin{table}[t]
\small
\caption{Kruskal--Wallis comparisons across the 10 repair groups (S = small, M = medium, L = large $\eta^2$).}
\label{tab:rq2_group_kruskal}
\centering
\begin{tabular}{lrrr}
\toprule
\textbf{Metric} & \textbf{$N$} & \textbf{$p$-value} & \textbf{$\eta^2$} \\ \midrule
Changed LOC & 197 & \textbf{0.0052} & \textbf{0.1198 (M)} \\
Changed files & 197 & \textbf{0.0014} & \textbf{0.1374 (M)} \\
Change entropy & 197 & \textbf{0.0034} & \textbf{0.1256 (M)} \\
Commits in fix & 197 & \textbf{0.0195} & \textbf{0.1008 (M)} \\
Completion latency & 188 & 0.4379 & 0.0481 (S) \\
Merged-PR duration & 133 & 0.5515 & 0.0593 (S) \\
Fixer experience (repo) & 173 & \textbf{0.0027} & \textbf{0.1468 (L)} \\
Fixer experience (global) & 173 & \textbf{0.0433} & \textbf{0.1010 (M)} \\ \bottomrule
\end{tabular}
\end{table}

\begin{table}[htbp]
\small
\caption{Significant post-hoc comparisons using Dunn's test with Bonferroni-adjusted $p$-values and directional $A_{12}$ (L = large).}
\label{tab:rq2_group_dunn}
\centering
\begin{tabular}{p{3cm}p{3.3cm}p{3.3cm}rr}
\toprule
\textbf{Metric} & \textbf{Group $a$} & \textbf{Group $b$} & \textbf{$p$} & \textbf{$A_{12}$} \\ \midrule
Changed LOC & Tool Schema/Discovery & Documentation & 0.0364 & 0.7654 (L) \\
Changed files & Tool Schema/Discovery & Documentation & 0.0238 & 0.7617 (L) \\
Changed files & Documentation & External Service & 0.0063 & 0.1936 (L) \\
Entropy & Startup/Environment & External Service & 0.0103 & 0.2153 (L) \\
Entropy & Documentation & External Service & 0.0018 & 0.1814 (L) \\
Commits & Startup/Environment & Tool Schema/Discovery & 0.0064 & 0.2716 (L) \\
Fixer experience (Repo) & Tool Execution & Documentation & 0.0365 & 0.7939 (L) \\ \bottomrule
\end{tabular}
\end{table}

\begin{figure}[!tb]
\centering
\subfloat[]{\includegraphics[width=0.48\textwidth]{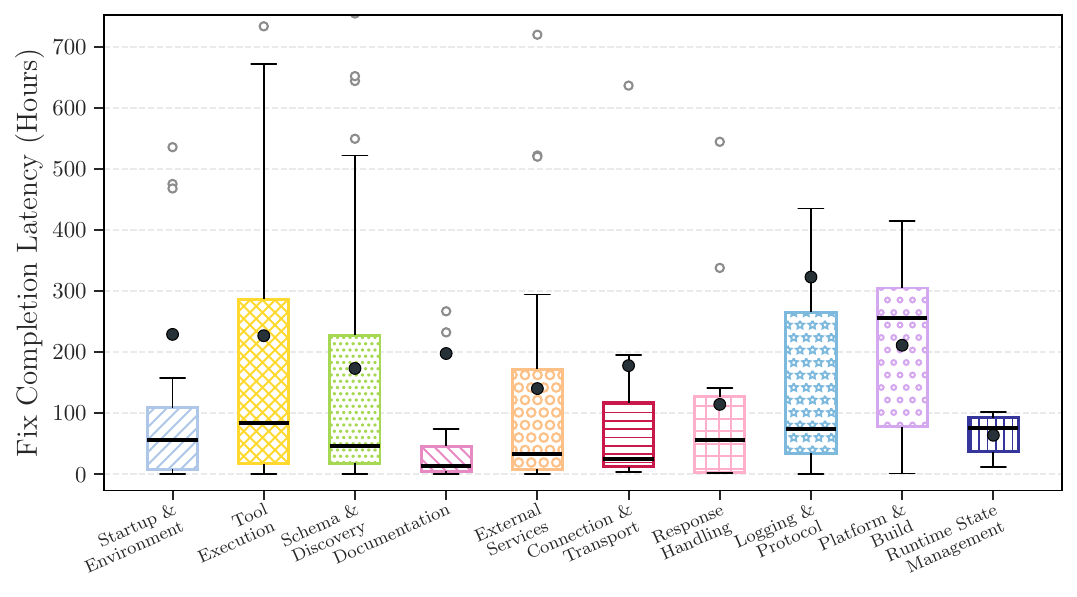}}
\subfloat[]{\includegraphics[width=0.48\textwidth]{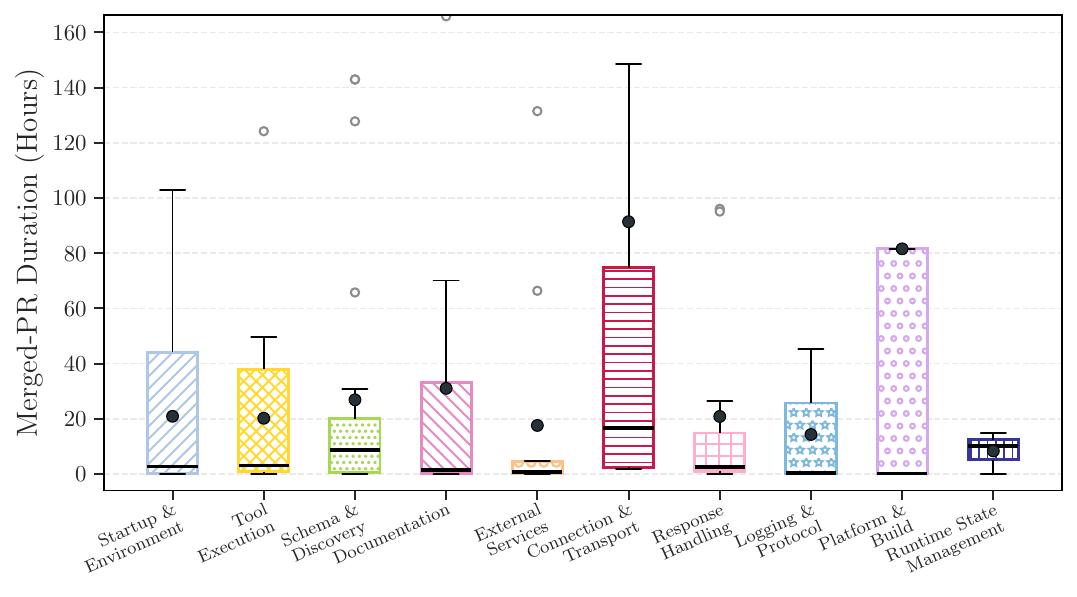}}\\
\subfloat[]{\includegraphics[width=0.48\textwidth]{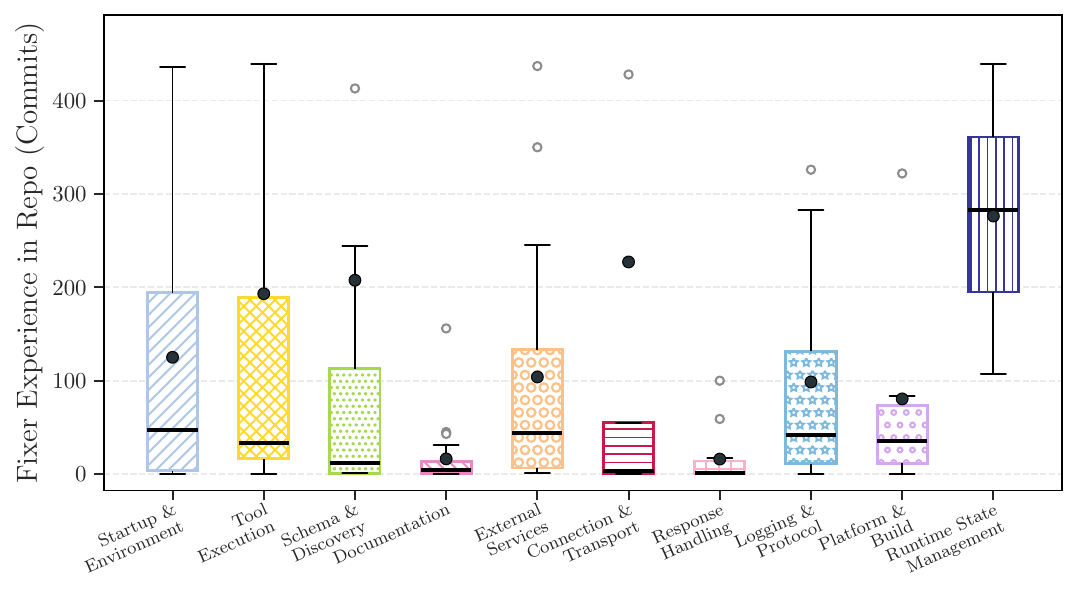}}
\subfloat[]{\includegraphics[width=0.48\textwidth]{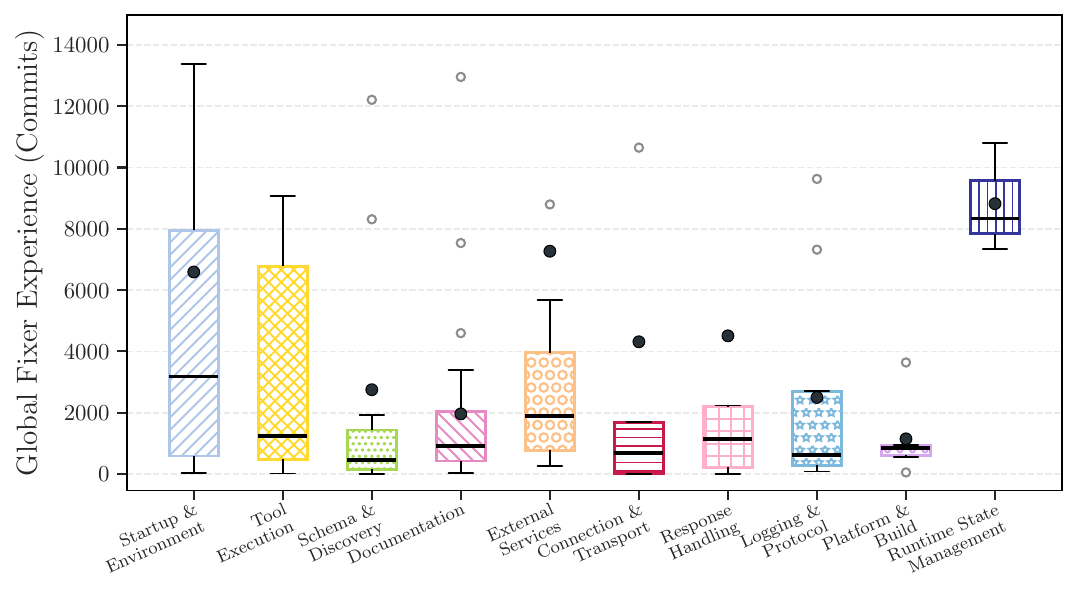}}
\caption{Process and developer-context metrics across the 10 repair groups. The panels show (a) non-negative fix completion latency ($N=188$), (b) merged-PR duration ($N=133$), (c) human fixer experience in the corresponding repository ($N=173$), and (d) global human fixer experience ($N=173$). Fixes without an applicable merged-PR duration are omitted only from panel (b), and bot-authored fixes only from panels (c) and (d).}
\Description{Four boxplots compare fix completion latency, merged pull-request duration, repository-level human fixer experience, and global human fixer experience across ten MCP repair groups. Completion latency and pull-request duration have no statistically detectable group difference; both experience measures differ overall, but only repository-level experience yields a significant post-hoc contrast.}
\label{fig:repair_group_process}
\end{figure}

\paragraph{Process and developer context.}
Using the metric-specific inclusion rules defined in Section~\ref{subsec:repair_metrics}, the repair-group comparisons include 188 repairs for completion latency, 133 for merged-PR duration, and 173 for each developer-context metric. Table~\ref{tab:rq2_process_medians} reports the group medians and applicable sample sizes. Figure~\ref{fig:repair_group_process} shows the distributions of the two process metrics and both developer-context metrics.

\begin{table}[!htb]
\footnotesize
\caption{Median process and developer-context characteristics across the repair groups. The $N$ columns report the applicable number of fixes within each group. Completion latency excludes fixes completed before issue creation; merged-PR duration includes only merged-PR fixes; experience excludes bot-authored fixes. Bold marks the maximum and unique minimum median in each metric column.}
\label{tab:rq2_process_medians}
\centering
\setlength{\tabcolsep}{2.4pt}
\begin{tabular}{lrrrrrrr}
\toprule
& \multicolumn{2}{c}{\textbf{Completion latency}} &
\multicolumn{2}{c}{\textbf{Merged-PR duration}} &
\multicolumn{3}{c}{\textbf{Human-fixer experience}} \\
\cmidrule(lr){2-3}\cmidrule(lr){4-5}\cmidrule(lr){6-8}
\textbf{Repair group} & \textbf{$N$} & \textbf{Median (h)} & \textbf{$N$} & \textbf{Median (h)} & \textbf{$N$} & \textbf{Repository} & \textbf{Global} \\ \midrule
Startup \& Environment & 40 & 56.45 & 25 & 2.71 & 36 & 47.0 & 3{,}178.0 \\
Tool Execution & 32 & 83.90 & 18 & 3.13 & 28 & 33.0 & 1{,}229.0 \\
Tool Schema \& Discovery & 28 & 46.54 & 25 & 8.70 & 27 & 12.0 & \textbf{459.0} \\
Documentation & 26 & \textbf{13.49} & 20 & 1.46 & 24 & 4.5 & 904.0 \\
External Service Integration & 19 & 32.43 & 12 & 0.80 & 15 & 44.0 & 1{,}903.0 \\
Connection \& Transport & 12 & 24.64 & 10 & \textbf{16.81} & 13 & 3.0 & 691.0 \\
Tool Response Handling & 12 & 56.01 & 11 & 2.46 & 13 & \textbf{1.0} & 1{,}140.0 \\
Logging \& Protocol Hygiene & 8 & 73.64 & 5 & 0.45 & 8 & 41.5 & 621.0 \\
Platform \& Build Compatibility & 6 & \textbf{255.53} & 4 & \textbf{0.16} & 6 & 35.5 & 847.5 \\
Runtime State Management & 5 & 75.15 & 3 & 10.18 & 3 & \textbf{283.0} & \textbf{8{,}338.0} \\ \bottomrule
\end{tabular}
\end{table}

Figures~\ref{fig:repair_group_process}(a) and~\ref{fig:repair_group_process}(b) show completion latency and merged-PR duration across the repair groups. Neither metric differs significantly across the groups according to the Kruskal--Wallis tests (Table~\ref{tab:rq2_group_kruskal}).
Figure~\ref{fig:repair_group_process}(c) and the post-hoc results reveal a narrower repository-experience difference. \emph{Tool Execution} fixes are associated with greater repository experience than \emph{Documentation} fixes, with a large $A_{12}$ effect. Executing and validating a tool may benefit from familiarity with repository-specific call paths, service abstractions, and tests, whereas a localized repair in \emph{Documentation} may be accessible with less project history. Figure~\ref{fig:repair_group_process}(d) shows that global fixer experience also varies widely; its Kruskal--Wallis test is significant, but no adjusted pairwise comparison remains significant. The evidence therefore points to local repository familiarity, rather than general GitHub contribution history, as the experience dimension that varies most clearly here.

\begin{keyfindings}[breakable=false]{RQ2}
\begin{keyfindingslist}
    \item Across the 221 fixes, the medians are two changed files and 59 changed LOC. Most repairs are localized, although a small number are much larger.
    \item Across metric families, changed LOC, changed files, and change entropy are positively correlated with completion latency ($\rho=0.22$--$0.26$). No structural metric is significantly correlated with merged-PR duration, and neither fixer-experience measure is significantly correlated with the structural or process metrics.
    \item Repair groups differ significantly in all four structural metrics. \emph{Tool Schema and Discovery} fixes change more LOC and files than \emph{Documentation} fixes and require more commits than \emph{Startup and Environment} fixes. \emph{External Service Integration} fixes modify more files than \emph{Documentation} fixes and have higher change entropy than \emph{Documentation} and \emph{Startup and Environment} fixes.
    \item Completion latency and merged-PR duration do not differ significantly across repair groups. The only significant post-hoc developer-context comparison shows greater repository experience for \emph{Tool Execution} than \emph{Documentation} fixes.
\end{keyfindingslist}
\end{keyfindings}
\subsection{RQ3: Construction and Composition of MCP-RepairBench}\label{subsec:rq3_benchmark}
RQ3 characterizes the tasks produced by the construction process and the evidence supporting the benchmark's use. We first report the benchmark's composition, behavioral coverage, and patch characteristics, and then present its validation and release outcome.

\subsubsection{Composition and coverage}
MCP-RepairBench contains 60 real-fault repair tasks from 25 repositories. Every task has a distinct issue and verified fix, and no two tasks share a buggy/fixed revision pair. The benchmark spans eight coverage groups representing the broad MCP behaviors addressed by the selected tasks (Table~\ref{tab:benchmark_composition_60}). The complete repository inventory appears in Table~\ref{tab:benchmark_repositories} in the appendix.

The distribution across the eight coverage groups reflects the purposive maximum-variation selection, which sought varied repair behaviors, while each selected case also had to support recoverable repository revisions, a clear task description, and repeatable offline evaluation. The relatively large number of \emph{Startup and Environment} tasks therefore describes the composition of MCP-RepairBench and does not estimate the frequency of these repairs in MCP repositories.

\begin{table}[h]
\small
\caption{Composition of MCP-RepairBench by benchmark coverage group.}
\label{tab:benchmark_composition_60}
\centering
\begin{tabular}{lr}
\toprule
\textbf{Coverage group} & \textbf{Cases} \\ \midrule
Startup \& Environment & 24 \\
Tool Execution & 14 \\
Tool Response Handling & 6 \\
Tool Schema \& Discovery & 6 \\
Logging \& Protocol Hygiene & 4 \\
Connection \& Transport & 3 \\
Documentation & 2 \\
External Service Integration & 1 \\ \midrule
\textbf{Total} & \textbf{60} \\ \bottomrule
\end{tabular}
\end{table}

Table~\ref{tab:benchmark_at_glance} characterizes the final tasks. The closed issues were created between December 16, 2024 and May 7, 2026, and task descriptions range from 23 to 1,706 words (median 170.5). Fifty-one tasks use a pull-request fix and nine use a direct commit. The largest repository contributes nine tasks (15\%), so repository dependence is limited but not eliminated.

\begin{table}[h]
\small
\caption{MCP-RepairBench at a glance.}
\label{tab:benchmark_at_glance}
\centering
\begin{tabular}{p{0.5\columnwidth}l}
\toprule
\textbf{Characteristic} & \textbf{Value} \\ \midrule
Tasks / repositories & 60 / 25 \\
Issue-creation date range & 2024-12-16 \textit{to} 2026-05-07 \\
Task-description words, median (mean; range) & 170.5 (258.5; 23--1,706) \\
Pull-request / direct-commit fixes & 51 / 9 \\
Runtime-behavior / configuration/package evaluators & 48 / 12 \\
Cases with selected pre-fix discussion & 9 \\
Cases with attachment transcriptions & 11 \\
Largest repository contribution & 9 (15\%) \\ \bottomrule
\end{tabular}
\end{table}

The complete developer fixes change a median of two files and 41 LOC, with maxima of 15 files and 2,994 LOC. After unrelated changes are excluded in nine cases (see Appendix~\ref{subsec:fix_test_scope} for an example), the issue-specific scopes have a median of two files and 37.5 LOC, with maxima of 10 files and 2,990 LOC. Of the complete fixes, 19 change one file and 34 change at most 50 LOC. The benchmark therefore contains both localized and substantially larger repairs.

\subsubsection{Release validation}
All 60 released tasks have buildable offline containers and benchmark-authored behavioral evaluators. Each evaluator produces the expected failure on the buggy state and success on the developer-fixed state; where applicable, preservation checks constrain incomplete repairs. Five offline executions of both states for every task yield 600/600 expected verdicts, supporting repeatability in the released environment.
\FloatBarrier

\begin{keyfindings}{RQ3}
\begin{keyfindingslist}
    \item MCP-RepairBench contains 60 real-fault repair tasks from 25 repositories, organized into eight benchmark coverage groups. It includes 48 runtime-behavior evaluators and 12 configuration/package evaluators.
    \item The developer fixes change a median of two files and 41 LOC, with maxima of 15 files and 2,994 LOC. The benchmark therefore includes both localized and substantially larger repairs.
    \item Five offline executions of the buggy and developer-fixed states for every task produce all 600 expected verdicts. Evaluators include preservation checks where applicable.
\end{keyfindingslist}
\end{keyfindings}

\subsection{RQ4: Effectiveness and Challenges of the MCP-Aware Repair Agent}\label{subsec:rq4_automation}
RQ4 evaluates the agent's repair effectiveness and API cost, then examines how results vary across comparison conditions and task subgroups. The analysis concludes with resource use, evaluator-feedback behavior, tool selection, and the recurring mechanisms behind unresolved tasks.

\subsubsection{Primary and repeated-trajectory results}
On MCP-RepairBench, the MCP-aware repair agent resolves 34 of 60 tasks on the first trajectory (56.7\%), which is the primary pass@1 result. A second fresh trajectory resolves five of the 26 remaining tasks, increasing the cumulative total to 39. A third trajectory resolves two of the remaining 21 tasks, producing a final pass@3 result of 41/60 (68.3\%) (Table~\ref{tab:rq4_agent_results}). Thus, fresh trajectories improve coverage, but the number of additional repairs falls from five on the second trajectory to two on the third. This observation is consistent with prior work, where the evaluation of an industrial repair agent similarly reported increased coverage across independent runs~\cite{engineeringAgent2026}. Kiecker et al. also observe diminishing returns under larger LLM repair-loop iteration budgets, although their study concerns additional iterations within a repair loop rather than fresh trajectories~\cite{kieckerRepairLoops2026}.

The 60 pass@1 trajectories cost \$4.253 in total, equivalent to \$0.071 per benchmark task and \$0.125 per successful repair. With up to three trajectories, early stopping results in 107 executed trajectories at a total cost of \$8.943. This corresponds to \$0.084 per executed trajectory, \$0.149 per benchmark task, and \$0.218 per successful repair. The incremental cost per additional repair rises from \$0.514 on the second trajectory to \$1.060 on the third.

\begin{table}[h]
\small
\caption{MCP-aware repair agent results on MCP-RepairBench. Pass@$k$ is the fraction resolved within at most $k$ independently initialized trajectories; early stopping avoids running later trajectories after success. Cost per repair is cumulative API cost divided by cumulative successful repairs.}
\label{tab:rq4_agent_results}
\centering
\begin{tabular}{lrrrrr}
\toprule
\textbf{Allowed trajectories} & \textbf{Newly solved} & \textbf{Cumulative} & \textbf{Rate} & \textbf{Cost} & \textbf{Cost/repair} \\ \midrule
First trajectory (pass@1) & 34 & 34/60 & 56.7\% & \$4.253 & \$0.125 \\
At most two trajectories (pass@2) & 5 & 39/60 & 65.0\% & \$6.823 & \$0.175 \\
At most three trajectories (pass@3) & 2 & 41/60 & 68.3\% & \$8.943 & \$0.218 \\ \bottomrule
\end{tabular}
\end{table}

\subsubsection{System and model comparisons}
\begin{figure}[htbp]
\centering
\subfloat[]{\includegraphics[width=0.85\textwidth]{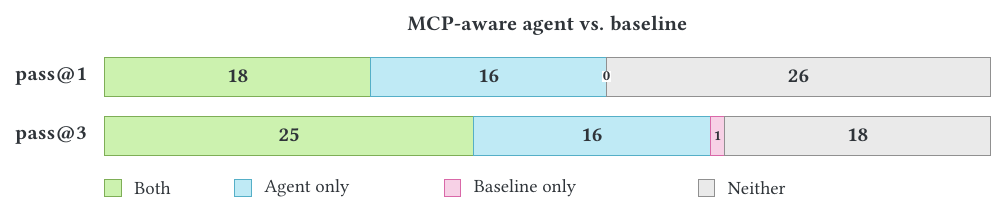}}\\[-2pt]
\subfloat[]{\includegraphics[width=0.85\textwidth]{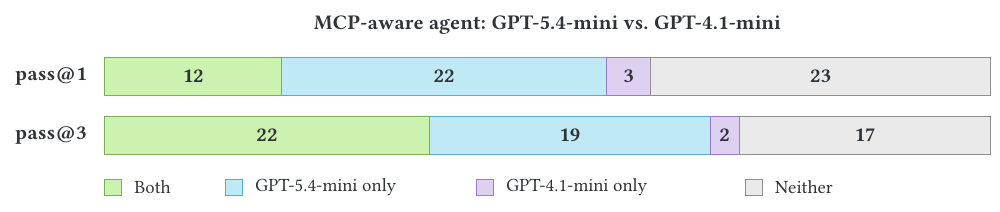}}
\caption{Task-level overlap in resolved outcomes at pass@1 and pass@3. Panel (a) compares the MCP-aware repair agent with the continuous-loop baseline. Panel (b) compares GPT-5.4-mini with GPT-4.1-mini within the MCP-aware agent. Each bar partitions the same 60 tasks according to whether both conditions, only one condition, or neither condition resolves them.}
\Description{Two stacked-bar panels compare task-level repair outcomes at pass@1 and pass@3. The first panel compares the MCP-aware repair agent with the baseline. The second compares GPT-5.4-mini with GPT-4.1-mini in the MCP-aware agent. Each bar partitions 60 tasks into those resolved by both conditions, by either condition alone, or by neither.}
\label{fig:rq4_paired_overlap}
\end{figure}

\paragraph{Baseline.}
Under the matched per-trajectory limits described in Section~\ref{subsec:agent_evaluation}, the baseline resolves 18 tasks at pass@1 and 26 at pass@3, compared with 34 and 41 for the MCP-aware agent (Table~\ref{tab:rq4_comparisons}).
The MCP-aware agent therefore resolves 16 more tasks at pass@1, a difference of 26.7 percentage points, and 15 more at pass@3, a difference of 25 percentage points.
Figure~\ref{fig:rq4_paired_overlap}(a) shows how the resolved tasks overlap between the two systems. Every task resolved by the baseline at pass@1 is also resolved by the MCP-aware agent. At pass@3, only one task is resolved by the baseline but not by the agent. The paired differences are statistically significant at both pass@1 and pass@3 (exact McNemar tests, $p<0.001$).

For the 16 tasks resolved only by the MCP-aware agent at pass@1, the baseline also produces a patch in every case. In 15 cases, both systems edit at least one common file. This suggests that the baseline usually identifies part of the relevant repair area but does not produce a complete passing repair. Inspection of the patches suggests that, compared with the baseline patches, the successful MCP-aware patches more often update all required files and interfaces or use implementations compatible with the repository's libraries and APIs.
On ten of these tasks, the MCP-aware
agent succeeds only after using evaluator feedback to revise an earlier candidate patch.

At pass@1, the baseline costs \$3.140 and the MCP-aware
agent costs \$4.253, but the agent has the lower cost per successful repair (\$0.125 versus \$0.174). At pass@3, early stopping results in 139 baseline trajectories and 107
MCP-aware agent trajectories. Their total costs are \$7.386 and \$8.943, respectively, corresponding to \$0.284 per baseline repair and \$0.218 per MCP-aware
agent repair. Thus, although the MCP-aware agent has a higher total API cost, its cost per successful repair is lower at both pass@1 and pass@3.

\paragraph{Model sensitivity.}
With GPT-4.1-mini, the MCP-aware repair agent resolves 15 tasks at pass@1 and 24 at pass@3, compared with 34 and 41 tasks, respectively, when using GPT-5.4-mini (Table~\ref{tab:rq4_comparisons}).
Figure~\ref{fig:rq4_paired_overlap}(b) shows the overlap between the tasks resolved with each model. At both pass@1 and pass@3, more tasks are resolved only with GPT-5.4-mini than only with GPT-4.1-mini. The paired differences are statistically significant at both pass@1 and pass@3 (exact McNemar tests, $p<0.001$).

Using each model's applicable token prices, the total pass@1 API cost is \$4.253 for GPT-5.4-mini and \$4.338 for GPT-4.1-mini. Although GPT-4.1-mini has lower uncached-input and output-token prices~\cite{openai_gpt54mini_model,openai_gpt41mini_model}, its lower repair success rate results in a higher cost per successful repair at pass@1 (\$0.289 versus \$0.125). An earlier exploratory baseline with GPT-4.1-mini resolved only 1/60 tasks despite more permissive resource limits. We therefore did not rerun this condition under the tighter matched limits.

\begin{table}[h]
\small
\caption{MCP-aware repair agent and continuous-loop baseline results from pass@1 through pass@3. The GPT-5.4-mini conditions use matched resource limits. Cost per repair is cumulative API cost divided by cumulative successful repairs.}
\label{tab:rq4_comparisons}
\centering
\begin{tabular}{lllrrrr}
\toprule
\textbf{Model} & \textbf{System} & \textbf{Pass} & \textbf{Solved} & \textbf{Rate} & \textbf{Cost} & \textbf{Cost/repair} \\ \midrule
\multirow{6}{*}{\texttt{gpt-5.4-mini}}
 & \multirow{3}{*}{MCP-aware agent} & pass@1 & 34/60 & 56.7\% & \$4.253 & \$0.125 \\
 & & pass@2 & 39/60 & 65.0\% & \$6.823 & \$0.175 \\
 & & pass@3 & 41/60 & 68.3\% & \$8.943 & \$0.218 \\ \cmidrule(lr){2-7}
 & \multirow{3}{*}{Continuous-loop baseline} & pass@1 & 18/60 & 30.0\% & \$3.140 & \$0.174 \\
 & & pass@2 & 23/60 & 38.3\% & \$5.547 & \$0.241 \\
 & & pass@3 & 26/60 & 43.3\% & \$7.386 & \$0.284 \\ \midrule
\multirow{3}{*}{\texttt{gpt-4.1-mini}}
 & \multirow{3}{*}{MCP-aware agent} & pass@1 & 15/60 & 25.0\% & \$4.338 & \$0.289 \\
 & & pass@2 & 21/60 & 35.0\% & \$8.102 & \$0.386 \\
 & & pass@3 & 24/60 & 40.0\% & \$11.507 & \$0.479 \\ \bottomrule
\end{tabular}
\end{table}

\subsubsection{Success relative to model knowledge cutoffs}
Because MCP-RepairBench uses public GitHub issues and fixes, some may have appeared in the evaluated models' training data. We compare each verified-fix date with the model's documented knowledge cutoff to examine whether the agent's successful repairs are limited to fixes dated before the cutoff. We also record whether the corresponding issue was created after the cutoff.
This comparison cannot establish whether an issue, fix, or related code was actually included in the LLM's training data. All 60 issues and verified fixes postdate GPT-4.1-mini's June~2024 cutoff~\cite{openai_gpt41mini_model}. Nevertheless, the MCP-aware agent with GPT-4.1-mini resolves 15 tasks at pass@1 and 24 at pass@3 (Table~\ref{tab:rq4_cutoff}). GPT-5.4-mini has a documented knowledge cutoff of August~2025~\cite{openai_gpt54mini_model}. Of the 60 tasks, 35 have verified fixes dated after the cutoff and 25 on or before it. The issues for 31 tasks were also opened after the cutoff. At pass@1, the MCP-aware agent with GPT-5.4-mini resolves 17 tasks in each fix-date group. At pass@3, these counts increase to 20 tasks with fixes on or before the cutoff and 21 with later fixes. The agent therefore succeeds on fixes dated both before and after the cutoff at pass@1 and pass@3.

\begin{table}[h]
\small
\caption{MCP-aware agent success by verified-fix chronology. Dates are compared with each model's documented knowledge cutoff using the fix artifact's creation or author date. The pass@1 and pass@3 columns report resolved tasks.}
\label{tab:rq4_cutoff}
\centering
\begin{tabular}{llrrr}
\toprule
\textbf{Model and cutoff} & \textbf{Verified-fix chronology} & \textbf{Tasks} & \textbf{pass@1} & \textbf{pass@3} \\ \midrule
\multirow{2}{*}{GPT-4.1-mini (2024-06-01)} & On/before cutoff & 0 & --- & --- \\
 & After cutoff & 60 & 15/60 (25.0\%) & 24/60 (40.0\%) \\ \midrule
\multirow{2}{*}{GPT-5.4-mini (2025-08-31)} & On/before cutoff & 25 & 17/25 (68.0\%) & 20/25 (80.0\%) \\
 & After cutoff & 35 & 17/35 (48.6\%) & 21/35 (60.0\%) \\ \bottomrule
\end{tabular}
\end{table}

\subsubsection{Subgroup performance}
Table~\ref{tab:rq4_groups} compares the MCP-aware agent and baseline by benchmark coverage group.
The largest difference occurs for \emph{Startup and Environment}. At pass@1, the agent resolves 11/24 tasks in this group, compared with 2/24 for the baseline; at pass@3, the corresponding results are 15/24 and 5/24. The agent also resolves all six \emph{Tool Schema and Discovery} tasks at pass@1, compared with three for the baseline. Results for groups containing only a few tasks are interpreted descriptively.

\begin{table}[t]
\small
\caption{MCP-aware repair agent and baseline success by benchmark coverage group at pass@1 and pass@3. Cells report resolved tasks over all tasks in the corresponding group.}
\label{tab:rq4_groups}
\centering
\begin{tabular}{lrrrr}
\toprule
& \multicolumn{2}{c}{\textbf{pass@1}} & \multicolumn{2}{c}{\textbf{pass@3}} \\
\cmidrule(lr){2-3}\cmidrule(lr){4-5}
\textbf{Coverage group} & \textbf{Agent} & \textbf{Baseline} & \textbf{Agent} & \textbf{Baseline} \\ \midrule
Connection \& Transport & 1/3 & 1/3 & 2/3 & 1/3 \\
Documentation & 2/2 & 2/2 & 2/2 & 2/2 \\
External Service Integration & 0/1 & 0/1 & 0/1 & 0/1 \\
Logging \& Protocol Hygiene & 2/4 & 1/4 & 2/4 & 2/4 \\
Startup \& Environment & 11/24 & 2/24 & 15/24 & 5/24 \\
Tool Execution & 10/14 & 7/14 & 11/14 & 9/14 \\
Tool Response Handling & 2/6 & 2/6 & 3/6 & 2/6 \\
Tool Schema \& Discovery & 6/6 & 3/6 & 6/6 & 5/6 \\ \midrule
\textbf{Total} & \textbf{34/60} & \textbf{18/60} & \textbf{41/60} & \textbf{26/60} \\ \bottomrule
\end{tabular}
\end{table}

Table~\ref{tab:rq4_evaluator_forms} reports success for the two evaluator forms. At pass@1, the agent resolves 32/48 tasks with runtime-behavior evaluators and 2/12 tasks with configuration/package evaluators. By pass@3, the corresponding totals increase to 36/48 and 5/12. No additional runtime-behavior task is resolved on the third trajectory, while both repairs added at pass@3 are configuration/package tasks. Repeated trajectories thus narrow the difference between the evaluator forms, but success remains lower for configuration/package tasks. This result points to the need for stronger inspection and validation of configuration/package artifacts during repair.

\begin{table}[t]
\small
\caption{Success by evaluator form. Pass@2 and pass@3 are cumulative under early stopping.}
\label{tab:rq4_evaluator_forms}
\centering
\begin{tabular}{lrrrrrr}
\toprule
& \multicolumn{2}{c}{\textbf{pass@1}} & \multicolumn{2}{c}{\textbf{pass@2}} & \multicolumn{2}{c}{\textbf{pass@3}} \\
\cmidrule(lr){2-3}\cmidrule(lr){4-5}\cmidrule(lr){6-7}
\textbf{Evaluator form} & \textbf{Solved} & \textbf{Rate} & \textbf{Solved} & \textbf{Rate} & \textbf{Solved} & \textbf{Rate} \\ \midrule
Runtime behavior & 32/48 & 66.7\% & 36/48 & 75.0\% & 36/48 & 75.0\% \\
Configuration/package & 2/12 & 16.7\% & 3/12 & 25.0\% & 5/12 & 41.7\% \\ \midrule
Total & 34/60 & 56.7\% & 39/60 & 65.0\% & 41/60 & 68.3\% \\ \bottomrule
\end{tabular}
\end{table}

\FloatBarrier
\subsubsection{Resource use, evaluator feedback, and tool use}
\begin{figure}[t]
\centering
\includegraphics[width=0.9\textwidth]{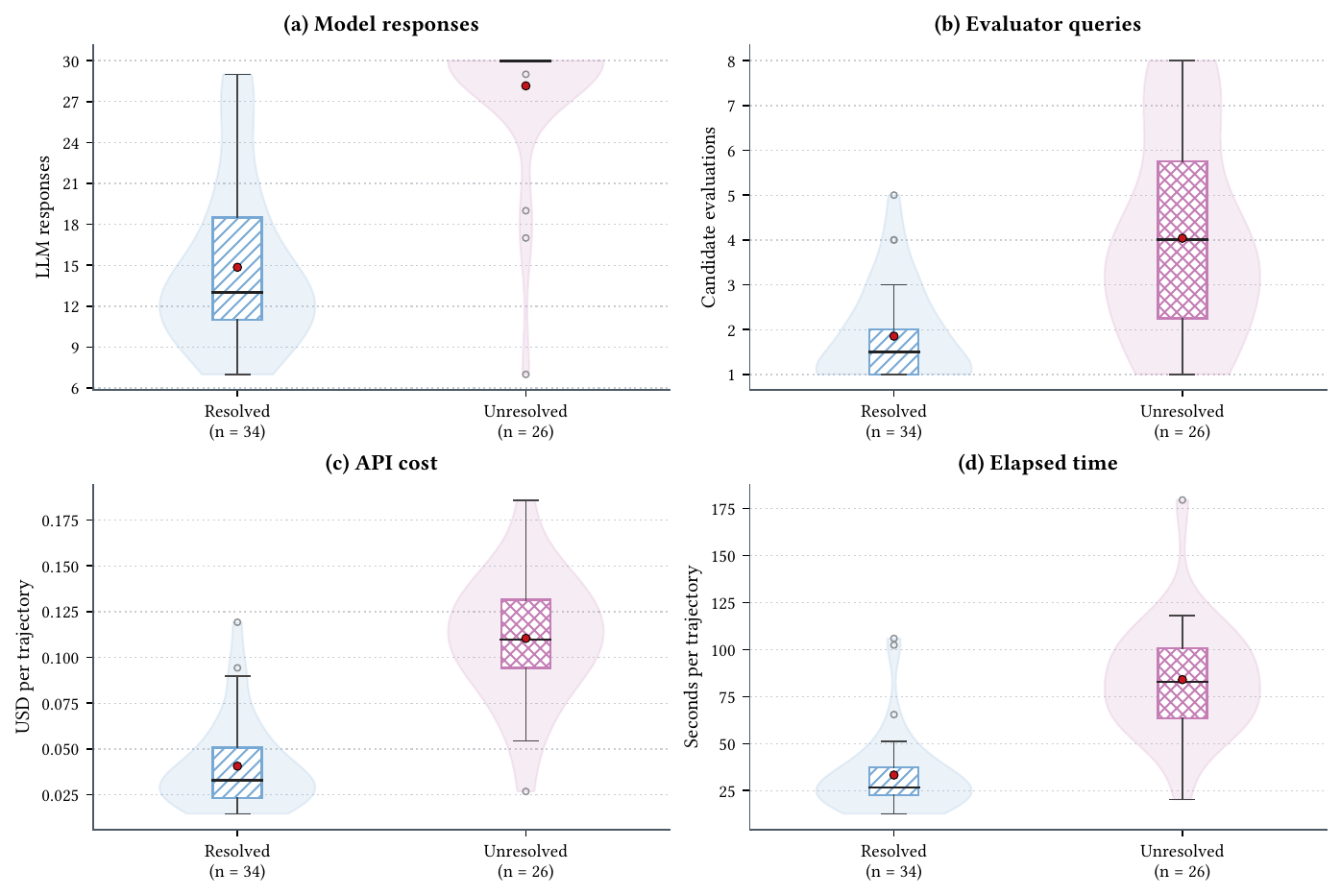}
\caption{First-trajectory resource distributions for 34 resolved and 26 unresolved tasks. Violin width shows density, open grey circles show outliers, and red circles show means. Elapsed time is measured from trajectory start to completion.}
\Description{Four violin and embedded box plots compare resolved and unresolved trajectories for model responses, evaluator queries, API cost, and elapsed seconds. Unresolved trajectories are concentrated near the response limit and are higher on all four measures.}
\label{fig:rq4_efficiency}
\end{figure}

We examine the first trajectories of the 34 resolved and 26 unresolved tasks to understand how resource use and agent activity differ by outcome. Figure~\ref{fig:rq4_efficiency} shows that unresolved trajectories use more model responses and evaluator queries and have higher API costs and elapsed times. The distributions of all four measures differ significantly between resolved and unresolved trajectories (two-sided Mann--Whitney U tests with Bonferroni correction, all adjusted \(p<0.001\)). The median number of model responses is 30 for unresolved trajectories and 13 for resolved trajectories. These differences partly reflect the stopping rule. A resolved trajectory ends when a candidate passes, whereas an unresolved trajectory may continue until it reaches a limit.

The MCP-aware repair agent produces a nonempty patch in 24 of the 26 unresolved first trajectories. Failure therefore usually occurs after editing has begun, rather than because the agent never produces a candidate repair.

Of the 34 resolved trajectories, 17 pass on the first evaluator query and nine on the second. The remaining eight pass between the third and fifth queries. Half of the successful trajectories therefore use feedback from at least one failed candidate before producing a passing repair. This pattern is consistent with industrial evidence that execution feedback can help repair agents revise candidate patches~\cite{engineeringAgent2026}.

\begin{figure}[t]
\centering
\includegraphics[width=0.9\textwidth]{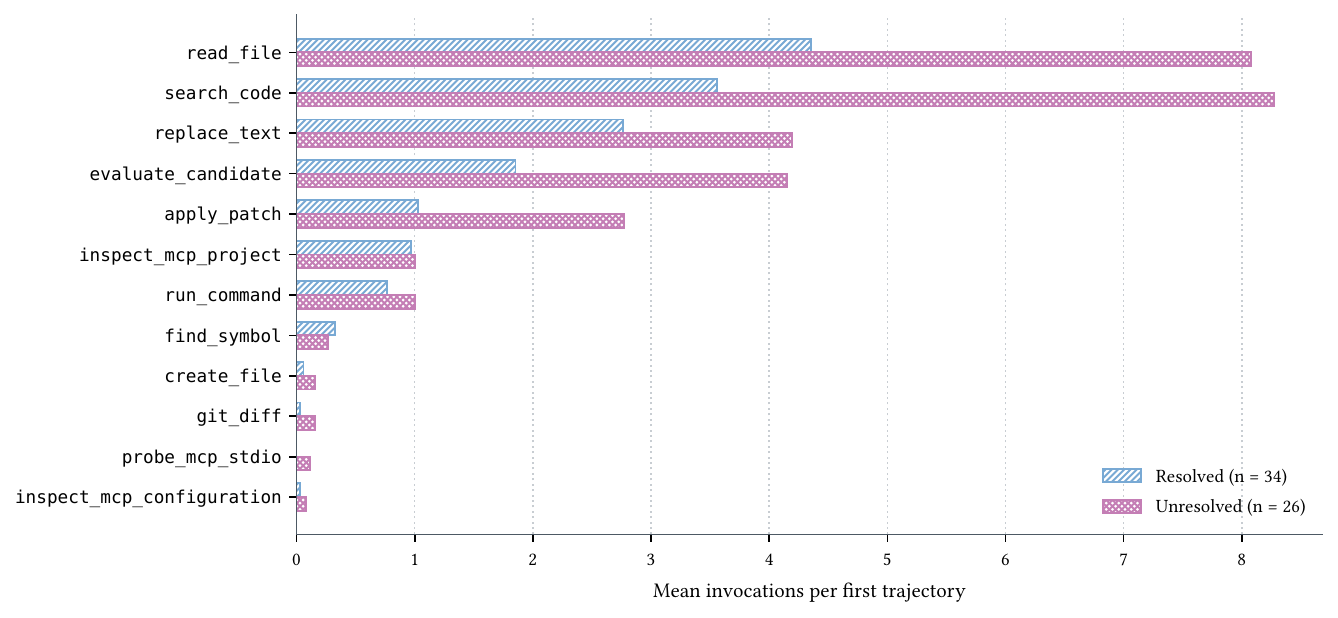}
\caption{Mean tool calls per first trajectory for 34 resolved and 26 unresolved tasks. Each mean includes trajectories in which the tool is not called. \texttt{probe\_mcp\_http} is omitted because it is never invoked.}
\Description{Grouped horizontal bars show mean invocations of repository, editing, evaluation, and MCP-specific tools for solved and unresolved trajectories. Unresolved cases use more generic inspection, editing, and evaluation calls.}
\label{fig:rq4_tools_used}
\end{figure}

Figure~\ref{fig:rq4_tools_used} reports the mean number of calls to each tool across the 34 resolved and 26 unresolved first trajectories. The raw means are generally higher for unresolved trajectories because these trajectories contain more model responses. We therefore also compare tool calls per ten model responses. Total normalized tool use is similar, at 10.50 calls for resolved tasks and 10.64 for unresolved tasks. In the observed traces, unresolved trajectories make relatively more calls to \texttt{search\_code} (2.92 versus 2.35) and \texttt{apply\_patch} (0.92 versus 0.56), but relatively fewer calls to \texttt{run\_command}, which executes local tests, builds, and diagnostic programs (0.35 versus 0.54).

Across all 60 first trajectories, \texttt{inspect\_mcp\_project}, which examines server construction, tool registration, transports, configuration, and output streams, is invoked in 59 tasks. In contrast, \texttt{inspect\_mcp\_configuration}, which traces a setting across environment variables, command-line options, defaults, and files, and \texttt{probe\_mcp\_stdio}, which exercises MCP interactions over stdio, are each invoked in only three tasks. \texttt{probe\_mcp\_http}, which checks a locally running MCP HTTP endpoint, is never invoked. The agent therefore uses general MCP project inspection routinely but rarely selects the specialized probes. Providing a specialized tool does not ensure that the model will recognize when to use it.

\subsubsection{Recurring failure mechanisms}
Nineteen tasks remain unresolved after three GPT-5.4-mini trajectories,
yielding 57 unsuccessful trajectories across these tasks. Of these trajectories, 55 produce a nonempty patch and 49 reach the 30-response limit. Only four reach the evaluator-query limit, and none reaches the API-cost or per-response output-token limit. The unsuccessful trajectories therefore usually end at the response limit after one or more candidate repairs have failed.

Inspection of task metadata, evaluator requirements, and execution traces identifies four recurring mechanisms.

\textbf{Overlooked configuration interfaces.}
Bugs involving server settings or declared dependency versions may require coordinated changes in server code and related configuration, package, container, or launch files. In one nearly complete repair, the agent added a command-line option for a custom browser path, read the same setting from an environment variable, and passed it to the browser. However, it represented the new field in the server's \texttt{BrowserConfig} class as a string rather than the required path type, leaving the configuration incomplete.\footnote{\url{https://github.com/stickerdaniel/linkedin-mcp-server/issues/118}}

\textbf{Incomplete error and output handling.} Some candidate patches correct the reported symptom but omit another required behavior. For example, one patch makes a failed Server-Sent Events (SSE) endpoint return HTTP status 500 but leaves the response body empty instead of returning the required \texttt{``Endpoint error''} message.\footnote{\url{https://github.com/baserow/baserow/issues/3709}} Another patch prevents routine status messages from being written to \texttt{stdout}, but also removes the error message shown when the required \texttt{truststore} library is unavailable. In both cases, the patch corrects one output while incorrectly changing or omitting another output checked by the evaluator.

\textbf{Removing the symptom instead of repairing the behavior.} Some candidate patches avoid the reported failure by disabling the operation involved rather than correcting it. For example, in one stdio case, a candidate revision removed the background function that reads from \texttt{stdin}, but did not replace it with the required check for whether the process that launched the server had ended.\footnote{\url{https://github.com/sooperset/mcp-atlassian/issues/935}} This stopped the competing read from \texttt{stdin}, but left the server unable to shut down automatically when that process ended.

\textbf{Inconsistent changes across files.} Some repairs require the same setting or version to be updated consistently in several files. For example, one candidate patch changed the required Python version in the continuous-integration (CI) workflow, Dockerfile, project metadata, and Renovate configuration, but left \texttt{.python-version} unchanged.\footnote{\url{https://github.com/stickerdaniel/linkedin-mcp-server/issues/389}} The repository therefore still declared conflicting Python versions. Across these patterns, the agent often finds the relevant code and makes a plausible edit, but leaves part of the required behavior unchanged or introduces an incompatibility. These failures show the need to identify all affected files and behaviors, use the mechanism required by the project, and verify both the corrected behavior and related existing behavior.

\begin{keyfindings}{RQ4}
\begin{keyfindingslist}
    \item The MCP-aware repair agent resolves 34/60 tasks at pass@1 and 41/60 at pass@3. The corresponding API costs are \$4.253 and \$8.943. The second and third trajectories add five and two repairs, respectively.
    \item With the same resource limits, the agent resolves 16 more tasks than the continuous-loop baseline at pass@1 and 15 more at pass@3. It also has a lower cost per successful repair. Results with GPT-4.1-mini show that repair success is sensitive to the underlying model.
    \item Most unsuccessful trajectories produce a candidate patch. The recurring failures involve overlooked configuration interfaces, incomplete error or output handling, removal of required behavior, and inconsistent changes across files.
\end{keyfindingslist}
\end{keyfindings}

\section{Discussion}
\label{sec:discuss}
This section integrates the four RQ findings. We interpret MCP repair as work across integration boundaries, derive implications for developers, SDK designers, repair systems, and benchmark design, and conclude by delimiting the claims supported by the two datasets and the evaluated MCP-aware agent configuration.

\subsection{MCP Repair as Integration-Boundary Repair}
The combined evidence supports a qualified interpretation. MCP repair frequently crosses integration boundaries. Taraghi et al. found that MCP faults concentrate at server configuration and server--tool or server--host boundaries~\cite{ourPaper}.
The repair groups in the present study reflect the same areas. \emph{Startup and Environment} and many \emph{Documentation} repairs concern server setup and host configuration, while \emph{Tool Execution} and \emph{Tool Schema and Discovery} concern interactions between servers and tools. Unresolved benchmark tasks also repeatedly involve configuration, dependencies, response handling, lifecycle behavior, and coordinated changes across files. Evidence from fault locations, developer repairs, and automated-repair failures therefore identifies these integration boundaries as recurring sources of maintenance work.%

An integration boundary in this study encompasses more than MCP message exchange. It includes the point at which a server is launched from a host configuration, a generated schema becomes a model-visible tool contract, an implementation value becomes an MCP response, a transport shares an output stream with application logging, or package metadata determines whether the server can start. A domain operation may be locally correct while one of these surrounding contracts remains inconsistent. Repair localization must therefore follow behavior across source code, configuration, packaging, documentation, and process lifecycle rather than assuming that the faulty line and the failed interface reside in the same artifact.

The 60 action patterns show that repairs within the same group can require different changes. RQ2 also finds significant differences among repair groups in changed LOC, changed files, change entropy, and commit count, but not in completion latency or merged-PR duration. Repair groups therefore help identify where repair work occurs, but they do not determine the exact change or the time required to complete it. Some repairs are confined to one file, while others require coordinated changes across several files or dependency updates.

\subsection{Implications for MCP Developers and SDK Designers}
The repair patterns suggest several opportunities to prevent boundary inconsistencies before release. Tool registration could validate names, required fields, array item definitions, optional-parameter semantics, and declared result types against the structures accepted by the implementation and MCP host. Tool response construction could similarly check structured content, error representation, and size limits rather than relying on successful serialization alone. These checks operationalize contracts already central to MCP tool discovery and invocation~\cite{mcpSpec2025}; placing them in SDKs or continuous integration could reduce repository-specific reinvention.

Servers should also validate defaults that affect startup and communication. A server using the stdio transport should write only MCP messages to \texttt{stdout} and send logs and error messages to \texttt{stderr}. Tests of stdio and HTTP communication should cover message formatting, required headers, configured endpoint paths, authentication, and clean shutdown. Startup checks should report whether each setting came from a command-line option, environment variable, configuration file, or default. They should also verify that required files and packaged resources exist and that dependency versions are compatible with both the server and any connected service. Packaging files, container definitions, and documented host configurations should be tested as parts of the server's public interface. Documentation warrants particular attention because installation-guidance clarifications and configuration-example corrections account for 19 of the 28 \emph{Documentation} repairs (Table~\ref{tab:rq1_action_distribution}). Users may copy these configurations directly, so an incomplete, malformed, or platform-specific example can prevent an otherwise functional server from starting.

Repairs involving initialization, cancellation, background operations, and shutdown also show that testing only successful initialization is insufficient. Tests should cover repeated initialization, cancellation of long-running requests, shutdown when a client or auxiliary process ends, and cases in which multiple components may attempt to read from \texttt{stdin} or wait for the same event. These tests cover behaviors that the evaluated agent sometimes failed to complete or preserve.

\subsection{Implications for Automated Repair}
For automated repair, this interpretation means that localization should extend beyond source files. Behavior at an integration boundary may be determined by dependency manifests and lockfiles, Dockerfiles, documented host configurations, code that reads environment variables, tool annotations and generated schemas, or code that initializes stdio and HTTP communication. Repair systems should identify the source and non-source artifacts that contribute to the reported behavior and follow the connections among them. For example, a documented environment variable should be traced to the code that reads it, the rules that determine its priority over other settings, and the command that launches the server. Similarly, a tool annotation should be traced to the generated schema, the function that handles the tool call, and the returned content. The lower success on configuration/package evaluators provides further evidence that repairs involving non-source artifacts remain challenging for the evaluated agent.

The evaluator feedback returned within the MCP-aware agent's repair loop should make the remaining failure clear. Useful feedback includes incorrect response fields, behavior under invalid inputs or service failures, and failed preservation checks. Of the 34 first-trajectory successes, 17 occur after at least one failed evaluator query, including ten of the 16 tasks resolved only by the MCP-aware agent. Candidate evaluation and revision are therefore common parts of successful trajectories. Similar repair systems also use execution feedback to revise candidate patches~\cite{repairagent2024,chatrepair2024}. The evaluator must still check the complete required behavior because incomplete tests can accept incorrect patches~\cite{qi2015,swebenchCorrectness2026}.

Specialized tools can contribute only when the model chooses them. The general MCP inspector was used broadly, but the configuration and stdio probes were rarely used, and the HTTP probe was never used. Future agents could recommend a specific probe when the inspected files or evaluator output match its purpose. Ablation studies could then measure how much these probes contribute to repair success.

The resource analysis shows that 24 of the 26 unresolved first trajectories produce a candidate patch. 
Their median response count reaches the 30-response limit, compared with a median of 13 responses for resolved trajectories.
Many failures therefore occur after editing has begun. Future agents could record which required behaviors each candidate addresses and compare the failures returned for successive candidates. If the same failure continues, the agent could inspect a different part of the repository or reconsider its diagnosis. It could stop when further edits are not supported by new evidence. These alternatives should be compared with the current stopping rule.

Fresh trajectories provide a complementary recovery mechanism, but their declining gains require balancing additional attempts against their cost. Future systems could use evidence from earlier failures to decide which unresolved tasks warrant another attempt rather than applying the same retry limit to every task.

\subsection{Implications for Benchmark Design}
Benchmark evaluators should check all behavior required by the issue and any existing behavior that a repair may affect. In RQ4, some candidate patches correct one output while omitting another, disable a failing operation instead of replacing it, or update only some of the files containing a required setting. These examples show why evaluators should include preservation checks and checks for plausible incomplete repairs while allowing implementations that differ from the developer patch. Each evaluator assertion should also be traceable to the issue and fix evidence~\cite{qi2015,swebenchCorrectness2026}.

Benchmarks built from issue reports measure problem diagnosis and patch generation together. MCP-RepairBench supplies pre-fix issue evidence rather than a complete benchmark-written specification of the required repair. The model must therefore infer the required behavior before implementing it. Scores should consequently be interpreted as end-to-end issue resolution, not patch generation from a complete specification~\cite{swebench2024}.

\subsection{Scope of the RQ4 Results}
The RQ4 results describe the two complete systems on the 60 MCP-RepairBench tasks. The MCP-aware agent and baseline share the model, reasoning setting, task descriptions, buggy repositories, and per-trajectory limits, but differ in their prompts, controllers, tools, and evaluator feedback. The comparison therefore does not isolate the effect of one component. Comparisons with results from other repair benchmarks would require aligned tasks, models, resource limits, evaluator access, and success criteria.

\section{Related Work}
\label{sec:related_work}
We position the study against four areas of related work. These are empirical fix patterns, MCP studies and benchmarks, automated program repair and repository-level agents, and repair benchmarks and evaluation validity.

\subsection{Empirical Fix Patterns}
Empirical repair studies commonly analyze repository history to identify recurring changes in human-written fixes. Pan et al. defined 27 automatically detectable patterns from the syntax and context of source-code hunk pairs and measured them across seven Java projects~\cite{panFixPatterns2009}. Their analysis focused on small changes and excluded hunk pairs containing more than seven statement lines. Islam and Zibran analyzed 4,653 bug-fixing revisions from five Java systems using AST differencing. They identified 38 fine-grained edit patterns and 37 nesting patterns that describe the code structures surrounding those edits~\cite{islamFixPatterns2020}.
Campos and Maia combined large-scale analysis of known bug-fix patterns with automated discovery of finer-grained repair actions~\cite{camposFixPatterns2019}. They measured five previously identified patterns in more than four million bug-fix commits from 101,471 Java projects. In a separate analysis of 395 Defects4J fixes from six projects, they extracted 155 fine-grained repair actions and identified 10 frequent cross-project actions.
Together, these studies show that source-code changes recur across projects. However, they represent repairs through source-code hunks, AST edits, and their syntactic contexts. Our RQ1 uses a different unit of analysis. It classifies verified issue--repair records according to the operation used to resolve the reported problem rather than the individual syntax-level edits in the patch.

FixMiner makes fine-grained source-code patterns usable for automated program repair~\cite{fixminer2020}. It analyzed 11,416 patches from 44 Java projects and grouped similar changes based on their code structure, edit actions, and tokens. Its repair prototype, PARFixMiner, generated correct patches for 26 of the 357 Defects4J bugs evaluated. FixMiner mines patterns for generating patches, whereas our taxonomy describes recurring repair operations in MCP software.

Studies of particular languages and domains show that general fix-pattern taxonomies may not capture language- and domain-specific repairs. Yang et al. identified Python-specific fix patterns that reflect features of the language, including dynamic typing~\cite{yangPythonFixPatterns2022}. Islam et al. provide a closer methodological precedent for our study. They used open coding to classify 415 Stack Overflow repairs and 555 GitHub repairs involving five Deep Learning (DL) libraries~\cite{islamDNNRepair2020}. Their classification included domain-specific operations such as repairing data dimensions and neural-network connectivity. They also examined repair challenges and produced a benchmark of 667 bug--repair instances.
Similarly, our RQ1 analysis develops a semantic repair taxonomy for MCP because general syntax fix patterns do not capture MCP-specific behavior. 

\subsection{MCP Studies and Benchmarks}
Research on MCP software is still developing. A broad ecosystem measurement found rapid adoption together with uneven project quality, maintenance, dependency, and client-connection characteristics~\cite{mcpMeasurement2025}. Taraghi et al. analyzed real MCP-server faults, derived five high-level fault categories, and found that faults frequently occur at component boundaries~\cite{ourPaper}. The present study builds on that work by examining how developers repair MCP faults.

Other MCP benchmarks primarily evaluate an agent's ability to use functioning servers and tools. Wang et al.'s MCP-Bench connects agents to 28 live MCP servers exposing 250 tools and evaluates tool discovery, parameter use, multi-step planning, cross-tool coordination, and task completion~\cite{wangMCPBench2026}. MCP-Universe exercises realistic tasks against multiple real-world MCP servers, while LiveMCP-101 stresses multi-step tool orchestration under changing external state~\cite{mcpUniverse2025,liveMCP1012025}. These benchmarks provide evidence about MCP-mediated tool use. MCP-RepairBench instead makes defective MCP software the evaluation object. A repair system receives an issue-derived problem description and a buggy repository and must produce a behaviorally valid patch.

\subsection{Automated Program Repair (APR) and Repository Agents}
Classical APR systems differ principally in how they define and search a repair space. GenProg searches mutations assembled from existing program material~\cite{genprog2009}; PAR instantiates manually designed patterns learned from human patches~\cite{par2013}; and TBar systematically applies a set of fix templates~\cite{tbar2019}. These techniques demonstrate how human repair knowledge can guide automation, while analyses of generated patches show that test-suite success alone can admit semantically incorrect repairs~\cite{qi2015}.

LLM-based repair approaches generate patches from code and natural-language context rather than applying only a fixed set of edit templates. Evaluations of code language models show their repair potential while also emphasizing the need for realistic tests and safeguards against training-data leakage~\cite{jiangCodeModelsAPR2023}. Tests can also guide the repair process rather than only assess the final patch. ChatRepair returns test failures to the model through a conversation, allowing it to revise candidate patches over multiple turns~\cite{chatrepair2024}.

Repository-level repair systems differ in how they gather context, control model interactions, and validate patches. RepairAgent~\cite{repairagent2024} allows an LLM to alternate between repository inspection, editing, and validation through repair-oriented tools, while a finite-state controller restricts tool use. SWE-agent~\cite{sweagent2024} focuses on the agent--computer interface and provides commands for repository navigation, editing, and execution. AutoCodeRover~\cite{autocoderover2024} uses issue text to retrieve relevant classes and methods before generating and validating a patch. MAGIS~\cite{magis2024} distributes issue resolution among agents with different roles. In an industrial setting, the Engineering Agent of Maddila et al.~\cite{engineeringAgent2026} starts from a triaged test failure, uses repository actions and analysis or test feedback, and submits validated patches to model-based and human review. Agentless~\cite{agentless2024} shows that localization, repair, and validation can also be organized as a structured pipeline without an open-ended agent loop. SpecRover~\cite{specrover2024} and SemAgent~\cite{semagent2025} instead emphasize inferred intent and semantic context.

Domain-specific systems provide closer precedents for MCP-aware evidence and actions. GradleFixer~\cite{gradleFixer2026} replaces a general shell with operations for inspecting and modifying Android/Gradle build environments. LLM-CompDroid~\cite{liu_llm_compdroid_2026} combines LLM generation with conventional analysis to repair Android XML compatibility faults. VulDebugger~\cite{vulDebugger2025} grounds C/C++ vulnerability repair in debugger states and inferred runtime constraints, while TerraRepair~\cite{terraRepair2026} retrieves Terraform dependencies and provider schemas and reruns infrastructure scanners before accepting a repair. Domain specialization can also be organized through cooperating roles. Smartify~\cite{smartify2025} assigns specialized agents to Solidity and Move vulnerability repair, VulnResolver~\cite{vulnResolver2026} combines repository context collection with safety-property analysis, and Alhanahnah et al.~\cite{alloyRepair2025} use a repair agent and an instructor agent to iteratively repair Alloy specifications from analyzer feedback. These systems specialize different parts of the evidence, action, or validation interface rather than relying only on domain wording in a prompt. Our agent follows the same general principle for MCP schemas, transports, processes, and configuration.

\subsection{Repair Benchmarks and Evaluation Validity}
Repair benchmarks encode different notions of a task and its oracle. ManyBugs and IntroClass~\cite{manybugsIntroClass2015} package reproducible C faults from mature systems and student programs, respectively, while Defects4J~\cite{defects4j2014} provides isolated Java faults with faulty and fixed revisions and tests that expose the faults. BEARS~\cite{bears2019} mines reproducible Java bug--fix pairs from continuous-integration history, and BugsInPy \cite{bugsinpy2020} adapts real fault screening and reproducible execution to Python. SWE-bench instead pairs a natural language GitHub issue with a full repository snapshot and evaluates a generated patch using fail-to-pass and pass-to-pass tests~\cite{swebench2024}. SWT-Bench uses a complementary issue-to-test formulation, asking systems to reproduce reported behavior before the developer fix is applied~\cite{swtbench2024}. Later SWE-bench variants expand the repair formulation through visual issue evidence, multiple programming languages, recent issues, automated environments, repeated execution, and expert review of issue sufficiency and test appropriateness~\cite{swebenchmultimodal2025,multiswebench2025,swebenchlive2025,swebenchverified2024}.

These benchmarks make system comparisons possible, but their scores remain conditional on task construction, included artifacts, and oracle strength~\cite{hirschDebugBenchmarks2022}. Reviews of modern APR and code benchmarks additionally identify inconsistent evaluation designs, contamination risks, and incomplete reporting or release practices~\cite{renzulloBenchmarkProblems2025,caoCodeBenchmarks2026}. Recent differential testing and manual inspection of plausible SWE-bench verified patches found that some test-passing patches diverge from intended behavior, reinforcing the long-standing plausible-versus-correct distinction~\cite{swebenchCorrectness2026,qi2015}. MCP-RepairBench addresses this risk through issue-specific behavioral checks and preservation checks rather than source equality with the developer patch. This improves observability for protocol, lifecycle, configuration, packaging, and documentation behavior, but it does not establish correctness beyond the encoded contract.

Taken together, prior work provides repair abstractions, repository-level workflows, and benchmark formulations, but does not connect real MCP fixes to an MCP repair benchmark and a controlled system evaluation. This paper fills that specific gap while keeping the three products analytically distinct. RQ1--RQ2 characterize repair evidence, RQ3 establishes benchmark composition and quality, and RQ4 uses the benchmark to identify attainable automation and remaining failure mechanisms.

\section{Threats to Validity}
\label{sec:threat}
We discuss threats to construct, internal, external, conclusion, and benchmark validity, together with the measures used to reduce them.\label{sec:threats}

\subsection{Construct Validity}
Linked PRs or commits may contain unrelated work, omit part of a repair, or fail to implement the reported fix. We reduce this risk by examining issue cross-references, discussions, diffs, and repository history, and by applying additional issue--fix and behavioral checks to benchmark cases. Multi-artifact fixes are combined when the artifacts jointly implement the repair. Undocumented or distributed repair work may nevertheless remain outside the selected artifacts.

The repair taxonomy also depends on human interpretation. Independent open coding of the \emph{What} and \emph{Why} descriptions, batch reconciliation, adjudication, affinity diagramming, and a final codebook recheck reduce individual coding error. Because the codes changed during open coding, we do not report agreement against a stable codebook. Excluding 24 large or diffuse fixes avoids assigning speculative action patterns, but RQ1 frequencies and RQ2 group comparisons consequently describe the 197 localizable repairs rather than all 221 fixes. Finally, changed LOC, changed files, commit count, change entropy, completion latency, PR duration, and prior-commit counts are proxies for patch structure, process, and developer experience. They do not directly measure effort, difficulty, or MCP expertise. We therefore define each measure explicitly, report complementary metrics, and avoid causal interpretations.

\subsection{Internal Validity}
The MCP-aware agent may revise candidates using output returned by the task evaluator. This feedback is part of the evaluated configuration and can improve outcomes relative to blind submission. Returned tracebacks can contain assertion names and internal evaluator paths. The continuous-loop comparison matches the model, reasoning effort, inputs, and per-trajectory limits, but the agent and the baseline differ in prompts, controllers, tools, and feedback. It therefore estimates the difference between complete systems, not the effect of an individual component. The two models also differ in reasoning support and other API behavior. Their comparison therefore measures sensitivity to the complete model condition rather than isolating the effect of knowledge cutoff or any other individual model property.

Phase~I repair evidence informed the general prompts and MCP-specific tool families, which may favor tasks with related characteristics. We mitigate direct leakage by constructing MCP-RepairBench from a separate issue pool. Benchmark developer patches, developer-fixed repository versions, evaluator files, and case-specific analyses were not used to develop the agent and remained unavailable to the repair model during evaluation. Each evaluation protocol and its per-trajectory limits were fixed before the corresponding experiments. Reported API costs exclude local computation, storage, and human development effort. Cross-model comparisons use each model's applicable token prices and therefore represent actual API expenditure rather than token-normalized efficiency.

\subsection{External Validity}
The datasets comprise open-source GitHub projects selected under Python-based eligibility criteria, and MCP libraries evolved during collection. Private deployments, other programming languages, different repository populations, and later protocol versions may exhibit different repair distributions. The extended corpus broadens repository and temporal coverage, but neither corpus is random. MCP-RepairBench contains 60 tasks selected for verified issue--fix evidence, offline evaluation feasibility, and coverage variation rather than prevalence, and some repositories contribute multiple tasks. We publish repository and coverage inventories and limit claims to the observed datasets and released tasks.

Evaluator images target Linux/ARM64. Repeated offline execution reduces environmental drift on that platform, but does not establish the same behavior on x86-64, other kernels, or different container runtimes.

\subsection{Conclusion Validity}
The repair metrics are non-normal and several repair groups are small. To this end, we use Spearman correlations for the cross-family analysis and Kruskal--Wallis and Dunn's tests for the repair-group comparisons. We use Mann--Whitney U tests for the four first-trajectory resource comparisons. We also apply Bonferroni correction where multiple relationships are tested and report effect sizes for the repair-group comparisons. We treat small-group and tool-use trace analyses as descriptive. Multiple cases from the same repository may also be dependent. These limitations restrict statistical generalization and do not support causal claims about group membership, experience, resource use, or individual agent components.

\subsection{Benchmark and Evaluator Validity}\label{benchmark_validity}
Benchmark-authored evaluators translate issue and fix evidence into behavioral contracts, and some replace credential-dependent or changing services with fixed local substitutes. These choices make evaluation reproducible and permit repairs that differ from the developer patch, but an evaluator may omit production behavior or reject an unanticipated valid repair. We check each evaluator on the buggy and developer-fixed states, add preservation and incomplete-repair checks, and execute both states repeatedly. Stage~A reviews task-description fidelity and solution leakage before Stage~B examines the complete construction evidence. One author conducted both stages, so this audit is weaker than independent expert review. A pre-release evaluator defect was corrected and the complete release gate rerun; undetected evaluator errors may still remain.

The LLM-assisted classification, summarization, clustering, and relevance screening used to find candidates can miss eligible tasks. Manual provenance review and validation protect the cases that reach construction, but cannot recover issues omitted earlier in the pipeline. MCP-RepairBench therefore supports reproducible evaluation of its released tasks, not complete coverage of MCP repairs.

Finally, public issue and fix histories make it impossible to guarantee that the benchmark was absent from model training data, a recognized concern for repair benchmarks~\cite{jiangCodeModelsAPR2023,renzulloBenchmarkProblems2025,ramosBugBenchmarkMemorization2025}. The agent was not developed from benchmark gold patches or failures. To assess possible cutoff effects, we compare every verified-fix date with the documented cutoff of each model used in the MCP-aware agent and report pass@1 and pass@3 separately for fixes dated on or before and after each cutoff (Table~\ref{tab:rq4_cutoff}). All fixes postdate GPT-4.1-mini's cutoff, while GPT-5.4-mini succeeds on fixes from both sides of its cutoff. The observed successes are therefore not limited to fixes dated on or before the models' documented cutoffs, although exposure to related code, repository conventions, or undocumented training data cannot be excluded.

\section{Conclusion}
\label{sec:conclusion}

This paper examined how developers repair MCP bugs and how real-world fixes can support the evaluation of automated program repair. We verified 221 fix-bearing issues and derived 10 repair groups and 60 action patterns from 197 localizable repairs. We then characterized the fixes using structural, process, and developer-context metrics.

The combined evidence shows that MCP repairs frequently cross integration boundaries. The largest repair groups are \emph{Startup and Environment}, \emph{Tool Execution}, \emph{Tool Schema and Discovery}, and \emph{Documentation}, while the 60 action patterns show that these repairs require diverse changes. Repair groups differ significantly in changed LOC, changed files, change entropy, and commit count, but not in completion latency or merged-PR duration.

We also constructed MCP-RepairBench from a separate extended corpus of bug reports. The benchmark contains 60 repair tasks from 25 repositories. On these tasks, the MCP-aware repair agent resolves 34 tasks on its first trajectory and 41 within three trajectories, compared with 18 and 26 tasks, respectively, for the baseline. Results with GPT-4.1-mini further show that repair success depends substantially on the underlying model. Unresolved tasks commonly involve exact response requirements, configuration or package changes, process lifecycle behavior, or incomplete changes across related files.

Future work can extend MCP-RepairBench across additional repositories, programming languages, platforms, protocol versions, and integration behaviors. Independent review of task descriptions and evaluators would further strengthen future benchmark releases. The MCP-aware repair agent could also be improved by refining its MCP-specific tools and directing the model to an appropriate probe when repository evidence or evaluator output indicates that specialized inspection is needed.

\bibliographystyle{ACM-Reference-Format}
\bibliography{_bibliography}

\appendix
\section{Supplementary Study and Benchmark Details}\label{appendix}
This appendix provides additional details about the benchmark tasks, evaluation records, review process, corpus composition, and RQ1 codebook.

\subsection{MCP-RepairBench Task and Reporting Protocol}

For each task, an evaluated system receives the approved issue-derived task description and a writable worktree, which is a modifiable copy of the repository at the buggy revision. It may inspect and edit that copy but cannot access the developer-fixed repository version, developer patch, analysis labels, review records, or evaluator files. For each evaluator query, the harness copies the current worktree to a temporary directory and evaluates that copy in a container with networking disabled. A task is resolved only if the evaluator exits successfully after checking the required behavior and any applicable preservation requirements. A benchmark report should identify the benchmark release, task inputs, repair system and model configuration, model sampling settings, trajectory and retry policy, resource limits, evaluator-feedback policy, case-level outcomes, and the number of resolved and attempted tasks.

Table~\ref{tab:benchmark_instance_fields} summarizes the components recorded for each released task. MCP-RepairBench follows the issue, repository, and evaluator structure used by repository-level repair benchmarks such as SWE-bench~\cite{swebench2024}. It also records the approved issue evidence, case review evidence, and offline repository assets.

\begin{table}[h]
\small
\caption{Core components of a released MCP-RepairBench task.}
\label{tab:benchmark_instance_fields}
\centering
\begin{tabular}{p{0.30\columnwidth}p{0.62\columnwidth}}
\toprule
\textbf{Component} & \textbf{Purpose} \\ \midrule
Provenance & Canonical repository, issue and fix URLs, and buggy and developer-fixed revisions \\
Task description & Curated pre-fix issue evidence without solution leakage \\
Evaluator & Behavioral checks and applicable preservation checks \\
Environment & Container definition, dependency evidence, and offline policy \\
Repository assets & Buggy and developer-fixed repository versions \\
Review record & Stage~A task-input decision and Stage~B case-acceptance decision \\
Public analysis fields & Benchmark coverage group, issue-specific changed files, and exclusion notes; hidden from the evaluated repair system \\
Retained construction evidence & Complete developer-fix scope, issue-specific scope, and developer diff; hidden from the evaluated repair system \\ \bottomrule
\end{tabular}
\end{table}

\subsection{Replication Packages and Agent-Evaluation Records}\label{subsec:artifact_ids}
The RQ1 and RQ2 empirical-study artifacts are available in the MCP repair study replication package~\cite{mcpRepairStudyArtifact}. The package contains the 197 coded repairs and associated codebooks, the 24 taxonomy exclusions, the canonical issue-level metrics for all 221 fix-bearing issues, provenance and exclusion records, analysis and plotting scripts, integrity tests, and the generated statistical results and figures.

The benchmark constructed in RQ3 and used in the RQ4 experiments is released as MCP-RepairBench \texttt{v1.0.0}~\cite{mcpRepairBenchArtifact}. Its replication package contains the 60 task descriptions, case metadata, review records, evaluator implementations, repository assets, container definitions, and scripts for preparing evaluation environments and assessing candidate repairs.

The evaluated repair systems and their experimental records are available in
the repair-agent package, release
\texttt{v1.0.1}~\cite{mcpRepairAgentArtifact}. The package contains the MCP-aware agent, continuous-loop baseline, prompts, experimental configurations, rerun harness, per-trajectory outcomes, generated patches, execution traces, tool-event records, token and cost accounting, and the case-level data underlying Figures~\ref{fig:rq4_efficiency} and~\ref{fig:rq4_tools_used}.

\subsection{Evaluator-Feedback Example}\label{subsec:evaluator_feedback_example}

Case \texttt{case\_0032\_linkedin\_docker\_python\_pin} illustrates the information returned after candidate evaluation. The first candidate changed how dependencies were installed but retained the Python~3.14 Docker image. The evaluator returned the following excerpt from the model-visible payload; generic traceback frames and fields not needed for the example are omitted.

\begin{lstlisting}[basicstyle=\ttfamily\scriptsize,breaklines=true,columns=fullflexible,frame=single]
{
  "query": 1, "passed": false, "exit_code": 1,
  "stderr": "... ReproError: Unexpected fixed observation:
    {\"uses_python_313_image\": false,
     \"uses_python_314_image\": true}",
  "instruction": "Use this concrete failure evidence to revise..."
}
\end{lstlisting}

The next candidate changed the Docker base image to Python~3.13. Its second evaluator query passed and reported that \texttt{uses\_python\_313\_image} was true and \texttt{uses\_python\_314\_image} was false. The sequence shows that evaluator output can reveal assertion names and observed values. The replication package contains the complete model-visible payload and trajectory~\cite{mcpRepairAgentArtifact}.

\subsection{Input Evidence Rules}

The task description may retain observed and expected behavior, reproduction steps, error messages, stack traces, protocol messages, relevant versions, deployment settings, and the pre-fix discussion needed to clarify the failure. It excludes links to the fixing PR or commit, patch excerpts, confirmed root causes, exact implementation instructions, post-fix confirmation, analysis labels, changed-file summaries, and validation hints derived from the developer fix. A filename that appears naturally in a stack trace is retained as evidence. A comment directing the repairer to change a named function is removed as a solution hint.

For the initial 60-candidate Stage~A review, automated temporal screening withholds 97 of 1,117 extracted issue and discussion segments because they occur after repair development begins. Human review then flags 19 assembled inputs for solution leakage, since a timestamp cutoff cannot remove a solution disclosed earlier in the issue history. Revised inputs are reviewed again before finalization. Across the released benchmark, nine tasks include 16 selected pre-fix discussion summaries and 11 tasks include 17 attachment transcriptions.

\subsection{Human Review Questions}

Stage A is restricted to the proposed task description and eligible pre-fix evidence. Its review form records an overall decision, whether additional issue-comment evidence is needed, and explanatory notes. The decision reflects whether the description (1) faithfully represents the report, (2) supplies enough evidence to attempt a repair, and (3) avoids root-cause or solution leakage. The reviewer may include, edit, or exclude the input and confirms every changed version before it is locked.

Stage B begins only after that lock. Its review form records five combined judgments:
\begin{enumerate}
    \item \emph{issue/fix validity}: the issue reports a real bug and the selected artifact implements the reported repair;
    \item \emph{contract fairness}: the required behavior and any preservation requirements follow from the evidence without prescribing an implementation;
    \item \emph{evaluator validity}: buggy failure and fixed success represent the stated behavior and reasonable alternative repairs can pass;
    \item \emph{preservation validity}: any preservation checks reject important regressions and superficial or incomplete repairs; and
    \item \emph{license adequacy}: the evidence permits the planned redistribution.
\end{enumerate}
The review record also contains the selected revisions, patch scopes, behavioral observations, duplicate or shared-fix evidence, environment evidence, and checks for incomplete repairs. A failed or unresolved judgment requires revision, exclusion, or replacement. Automated checks provide evidence but do not replace the human decision.

\subsection{Fix and Test Scope}\label{subsec:fix_test_scope}

The complete developer-fix scope includes every change in the linked PR or commit, including implementation, tests, documentation, lockfiles, and any unrelated work submitted with the repair. The issue-specific scope includes only changes that implement, test, configure, or directly document the repair for the selected issue. It retains repair-specific tests but can exclude unrelated ignore rules, release notes, generated files, features, or work for another report. Nine released tasks have a documented exclusion from the issue-specific scope. For example, the complete repair for \texttt{sooperset/mcp-atlassian} issue~\#179 changes \texttt{README.md} and \texttt{.gitignore}, while the issue-specific scope retains only the README correction.\footnote{The issue is \url{https://github.com/sooperset/mcp-atlassian/issues/179}, and the repair is \url{https://github.com/sooperset/mcp-atlassian/pull/197}.} These two scopes are used to characterize the developer fixes, not to score generated patches. The repair system does not receive the developer's test changes, and a generated patch need not modify the same files as the developer repair.

Every evaluator reproduces the behavior described by the issue. Forty-eight runtime-behavior evaluators execute affected project code with controlled inputs. Twelve configuration/package evaluators inspect configuration, dependency, packaging, Docker, or documentation files that determine the required behavior. Existing tests, issue examples, and the developer fix inform these checks. Where applicable, preservation checks and checks for incomplete repairs prevent deletion, hard-coding, or a change to only one affected interface from receiving credit for correcting a single observed symptom.

\subsection{Benchmark Candidate Replacements}

Sixteen of the 60 initially selected candidates were replaced at different points in the construction process. Before Stage~A, we checked that the issue and linked repair remained publicly accessible, that the buggy and developer-fixed revisions could be recovered, that the reported repair could be represented as one independent task, and that the candidate did not duplicate another repair. These checks before construction led to 11 replacements. Stage~A then examined only the proposed task description and eligible pre-fix evidence; three additional candidates could not support a sufficiently clear and non-leaking task based on one independent repair. Stage~B reviewed the completed candidate case and replaced two further candidates. One selected developer-fixed revision did not demonstrate the required behavior, and the other candidate required the same underlying repair as another case.

Table~\ref{tab:benchmark_exclusions} reports one primary reason for each replacement. Revisions to an otherwise admissible task description, including the leakage corrections reported under the input-evidence rules, are not counted as replacements. Every replacement candidate underwent the checks before construction and entered manual review at Stage~A. The released benchmark consequently contains 44 candidates from the initial selection and 16 replacements.

\begin{table}[h]
\small
\caption{Stage and primary reason for replacing 16 initially selected benchmark candidates.}
\label{tab:benchmark_exclusions}
\centering
\begin{tabular}{>{\raggedright\arraybackslash}p{0.19\columnwidth}>{\raggedright\arraybackslash}p{0.59\columnwidth}r}
\toprule
\textbf{Stage} & \textbf{Primary reason for replacement} & \textbf{Candidates} \\ \midrule
Before construction & Issue or linked-repair evidence no longer publicly accessible & 9 \\
Before construction & Multipurpose repair could not be isolated as one task & 1 \\
Before construction & Repair duplicated another candidate & 1 \\ \midrule
Stage~A & Issue and repair evidence did not support one defensible task & 1 \\
Stage~A & A sufficient non-leaking input and implementation-neutral evaluation could not both be constructed & 1 \\
Stage~A & Repair required multiple dependent stages that could not be represented as one task & 1 \\ \midrule
Stage~B & Selected developer-fixed revision did not demonstrate the required behavior & 1 \\
Stage~B & Candidate required the same underlying repair as another task & 1 \\ \midrule
\multicolumn{2}{r}{\textbf{Total replacements}} & \textbf{16} \\ \bottomrule
\end{tabular}
\end{table}
\subsection{Repository Composition}

\begin{table}[h]
\small
\caption{Final MCP-RepairBench tasks per canonical repository.}
\label{tab:benchmark_repositories}
\centering
\begin{tabular}{p{0.36\columnwidth}r p{0.36\columnwidth}r}
\toprule
\textbf{Repository} & \textbf{$N$} & \textbf{Repository} & \textbf{$N$} \\ \midrule
sooperset/mcp-atlassian & 9 & dbt-labs/dbt-mcp & 7 \\
neo4j-contrib/mcp-neo4j & 5 & stickerdaniel/linkedin-mcp-server & 5 \\
clickhouse/mcp-clickhouse & 4 & omarbenhamid/django-mcp-server & 3 \\
starrocks/mcp-server-starrocks & 3 & zilliztech/mcp-server-milvus & 3 \\
genomoncology/biomcp & 2 & pipeboard-co/meta-ads-mcp & 2 \\
saidsurucu/yargi-mcp & 2 & semgrep/mcp & 2 \\
alelec/pip-system-certs & 1 & andybrandt/mcp-simple-openai-assistant & 1 \\
apache/doris-mcp-server & 1 & baserow/baserow & 1 \\
dremio/dremio-mcp & 1 & gongrzhe/office-word-mcp-server & 1 \\
keboola/mcp-server & 1 & lpigeon/ros\_mcp\_server & 1 \\
oceanbase/awesome-oceanbase-mcp & 1 & oceanbase/mcp-oceanbase & 1 \\
qdrant/mcp-server-qdrant & 1 & robotmcp/ros-mcp-server & 1 \\
rohitg00/kubectl-mcp-server & 1 & & \\ \midrule
\multicolumn{3}{r}{\textbf{Total}} & \textbf{60} \\ \bottomrule
\end{tabular}
\end{table}
\FloatBarrier

\subsection{Benchmark Fix Artifact Types}

The released case metadata records the programming languages and file formats represented among the files modified by each verified developer fix. These labels describe the fix artifacts, not the repository's primary language or information supplied to the repair agent. Counts are non-exclusive because 20 of the 60 fixes modify more than one language or file-format category.

\begin{table}[h]
\small
\caption{Languages and file formats represented in the verified fixes. Counts are non-exclusive.}
\label{tab:benchmark_fix_artifact_types}
\centering
\begin{tabular}{lr@{\hspace{2.5em}}lr}
\toprule
\textbf{Type} & \textbf{Cases} & \textbf{Type} & \textbf{Cases} \\ \midrule
Python & 50 & TOML & 6 \\
Markdown & 11 & JSON & 4 \\
YAML & 9 & Dockerfile & 2 \\
TypeScript & 1 & JavaScript & 1 \\ \bottomrule
\end{tabular}
\end{table}
\FloatBarrier

\subsection{Release Verification Evidence}

MCP-RepairBench includes the 60 task descriptions, buggy and developer-fixed repository versions, evaluator implementations, review records, and offline execution environments. Each evaluator runs with networking disabled. Five executions on both repository versions for every task yield all 600 expected verdicts. Diagnostic details such as temporary paths, generated identifiers, or temporary port numbers may vary without changing whether the evaluator passes or fails.

\subsection{RQ1 Action-Pattern Codebook}\label{subsec:rq1_codebook}

The 10 repair groups identify the broad area addressed by a repair, while the 60 action patterns identify the change made by the developer. The table below provides the complete action-pattern codebook used for the 197 coded repairs. Counts are aggregated by action-pattern name. The two patterns observed in more than one group therefore list both groups in one row.

\scriptsize
\renewcommand{\arraystretch}{1.03}
\begin{longtable}{p{0.2\textwidth}p{0.23\textwidth}r p{0.45\textwidth}}
\caption{Complete RQ1 action-pattern codebook. Counts preserve the reviewed 197-issue coding.}\label{tab:rq1_codebook}\\
\toprule
\textbf{Repair group} & \textbf{Action pattern} & \textbf{$N$} & \textbf{Definition} \\ \midrule
\endfirsthead
\multicolumn{4}{c}{\tablename\ \thetable{} (continued)}\\
\toprule
\textbf{Repair group} & \textbf{Action pattern} & \textbf{$N$} & \textbf{Definition} \\ \midrule
\endhead
\midrule
\multicolumn{4}{r}{Continued on next page}\\
\endfoot
\bottomrule
\endlastfoot
Connection \& Transport & endpoint/path correction & 5 & Corrects incorrect connection, route, URI, or endpoint paths used between MCP components or connected systems. \\
Connection \& Transport & network concurrency handling & 3 & Coordinates or replaces transport-side scheduling and concurrency so long-running or platform-specific MCP communication remains responsive and connected. \\
Connection \& Transport & request header correction & 1 & Corrects the name, casing, or binding of a request header so the MCP transport or authentication flow recognizes the required session metadata. \\
Connection \& Transport & transport configuration & 2 & Adds or corrects server, host, or container settings needed to expose and use the intended MCP transport. \\
Connection \& Transport & transport protocol adaptation & 1 & Adapts transport message framing, wrapping, or session handling to the protocol behavior expected by the current MCP implementation. \\
Connection \& Transport; External Service Integration & DB connection correction & 3 & Corrects the construction, encoding, or selection of a database connection target so the server uses the intended database configuration. \\
Documentation & configuration example correction & 9 & Corrects broken or misleading sample configuration snippets used in MCP client or host setup. \\
Documentation & connection guidance clarification & 2 & Clarifies how to configure, establish, or confirm the connection between an MCP server and a host or client. \\
Documentation & feature availability clarification & 1 & Corrects documentation about where or under which conditions an optional MCP-related feature is supported. \\
Documentation & formatting correction & 2 & Repairs documentation layout, spacing, or embedded media so setup and usage guidance is readable and complete. \\
Documentation & installation guidance clarification & 10 & Clarifies how to install, launch, or configure the MCP server so users can get it running successfully. \\
Documentation & path correction & 2 & Corrects a path in documentation or documentation build configuration so referenced content or user configuration locations resolve correctly. \\
Documentation & security guidance addition & 1 & Adds missing security warnings or precautions for deploying or connecting to MCP capabilities with material trust or execution risks. \\
Documentation & transport configuration clarification & 1 & Corrects or clarifies documented transport options and launch instructions so users select a compatible MCP connection mode. \\
External Service Integration & concurrency handling & 1 & Bounds or coordinates concurrent external-service requests so tool execution does not exhaust connections or other provider resources. \\
External Service Integration & configuration parameter addition & 1 & Adds or changes a provider-facing configuration parameter or default needed for external-service operations to execute successfully. \\
External Service Integration & dependency version compatibility & 1 & Provides or selects dependency variants compatible with the supported versions of an external service or its client library. \\
External Service Integration & external API lookup fallback & 1 & Adds an alternative external-API lookup path when the primary lookup returns incomplete or empty results. \\
External Service Integration & provider configuration management & 7 & Corrects how the MCP server configures external providers such as credentials, regions, SSL, or provider-specific settings. \\
External Service Integration & request/endpoint correction & 5 & Corrects how the MCP server builds or routes requests to external-service endpoints during tool execution. \\
External Service Integration & transaction handling correction & 1 & Corrects external database transaction handling so write or schema-changing operations are committed and become observable. \\
Logging \& Protocol Hygiene & log level management & 4 & Adjusts log severity or suppresses non-critical logs so normal runtime output is not surfaced as an error. \\
Logging \& Protocol Hygiene & protocol output isolation & 4 & Separates non-protocol output from the MCP communication channel so JSON or stdio transport remains clean. \\
Platform \& Build Compatibility & container/build configuration correction & 3 & Corrects container or build configuration, dependency inclusion, health checks, or package metadata needed to build and start the server artifact. \\
Platform \& Build Compatibility & package path correction & 1 & Corrects the package or module path used by the build configuration so the installed artifact contains importable server code. \\
Platform \& Build Compatibility & platform support addition & 2 & Adds build, architecture, container, or host-network settings required to support an additional target platform. \\
Runtime State Management & runtime task synchronization & 1 & Serializes or coordinates competing initialization and tool-execution tasks so shared runtime services remain available and consistent. \\
Runtime State Management & server lifetime management & 1 & Adds or corrects startup, timeout, or shutdown control for an auxiliary server process so it terminates at the intended lifecycle boundary. \\
Runtime State Management & session state synchronization & 3 & Keeps configuration, authentication, project, or other mutable state consistent with the active MCP session, or removes session-local tracking that creates inconsistencies. \\
Startup \& Environment & dependency import correction & 3 & Updates an import path or package scope after dependency API or package-layout changes so the MCP server can start. \\
Startup \& Environment & dependency installation correction & 2 & Corrects dependency groups or installation steps so required runtime packages are installed without forcing unavailable optional dependencies. \\
Startup \& Environment & dependency replacement & 2 & Removes or replaces a dependency whose build, compatibility, or runtime behavior prevents reliable MCP installation or execution. \\
Startup \& Environment & package export correction & 1 & Exports a required class or function from the documented package namespace so consumers can import the MCP integration successfully. \\
Startup \& Environment & platform compatibility support & 3 & Adapts startup detection, launch commands, or host configuration to work across supported operating systems or execution environments. \\
Startup \& Environment & Python compatibility correction & 3 & Corrects Python-version-specific annotations, hooks, interpreter selection, or runtime behavior so installation and startup work on supported Python environments. \\
Startup \& Environment & startup configuration handling & 8 & Corrects how startup settings, environment variables, or configuration values are read, defaulted, or validated. \\
Startup \& Environment & startup initialization control & 1 & Controls one-time initialization work during startup so repeated setup does not interfere with server registration or readiness. \\
Startup \& Environment & startup path/resource setup & 8 & Corrects missing, wrong, or mislocated files, directories, packaged resources, or paths needed at startup. \\
Startup \& Environment & startup/runtime configuration correction & 5 & Corrects runtime startup wiring such as launch commands, app setup, route setup, or deployment configuration. \\
Startup \& Environment; Tool Schema \& Discovery & dependency version update & 9 & Updates a dependency version to restore compatibility or correct behavior affecting MCP startup or execution. \\
Tool Execution & execution environment adaptation & 4 & Adapts tool execution to the surrounding runtime or host environment so the tool can run correctly there. \\
Tool Execution & fallback/exception handling & 2 & Adds a fallback or handles an exceptional or missing runtime value so tool execution can recover or fail predictably instead of crashing. \\
Tool Execution & file I/O handling & 4 & Corrects reading, writing, encoding, or temporary-file behavior used during MCP tool execution or server operation. \\
Tool Execution & parameter handling correction & 4 & Corrects how execution-time parameters are accepted, passed through, or interpreted by the tool implementation. \\
Tool Execution & path handling & 4 & Corrects how filesystem paths are interpreted, normalized, escaped, checked, or applied during execution. \\
Tool Execution & request/command formatting correction & 4 & Corrects the format of outbound requests or commands so downstream systems receive valid inputs. \\
Tool Execution & resource state handling & 3 & Preserves, resolves, or normalizes persistent resource metadata and identity during tool operations so later reads and updates remain consistent. \\
Tool Execution & subprocess execution handling & 1 & Corrects subprocess input or invocation behavior required for a tool operation to complete in supported repository states. \\
Tool Execution & test coverage addition & 1 & Adds regression coverage for a previously failing tool operation so the behavior is reproducible and protected against recurrence. \\
Tool Execution & timeout/concurrency handling & 3 & Adjusts timeouts, long-running work, or concurrent execution so tool calls do not fail prematurely, hang, or consume excessive resources. \\
Tool Execution & tool input management & 2 & Corrects parsing or validation of tool inputs so valid arguments are accepted and interpreted without weakening necessary safeguards. \\
Tool Response Handling & tool response error handling & 4 & Corrects how tool failures or missing values are handled so MCP returns usable error information instead of crashing. \\
Tool Response Handling & tool response size reduction & 4 & Reduces oversized tool outputs so MCP responses remain usable within host and model context limits. \\
Tool Response Handling & tool response string format correction & 2 & Corrects string construction or quoting used to produce tool responses so the server parses and emits the response successfully. \\
Tool Response Handling & tool response structure correction & 1 & Preserves or corrects the organization and metadata of returned tool data so clients can interpret records and references accurately. \\
Tool Response Handling & tool response type alignment & 2 & Aligns returned values and content objects with declared MCP response types so response construction and client decoding succeed. \\
Tool Schema \& Discovery & tool description clarification & 1 & Adds missing argument-format or usage guidance to a tool description so models can construct valid invocations. \\
Tool Schema \& Discovery & tool name correction & 5 & Corrects invalid, ambiguous, or overly long tool names so hosts can accept and invoke them properly. \\
Tool Schema \& Discovery & tool registration correction & 7 & Corrects how tools are exposed or registered so they appear and function correctly in the MCP host. \\
Tool Schema \& Discovery & tool signature correction & 15 & Corrects the declared tool input structure so hosts and models can call the tool with the right argument types and shape. \\
\end{longtable}

\end{document}